\documentclass[sigconf,screen,authorversion]{acmart}
\AtBeginDocument{%
  \providecommand\BibTeX{{%
    \normalfont B\kern-0.5em{\scshape i\kern-0.25em b}\kern-0.8em\TeX}}}

\setcopyright{acmcopyright}
\copyrightyear{2024}
\acmYear{2024}
\acmDOI{10.1145/3640792.3675739}

\acmConference[AutoUI '24]{16th International ACM Conference on Automotive User Interfaces and Interactive Vehicular Applications}{September 22--25, 2024}{Stanford, CA}
\acmBooktitle{16th International ACM Conference on Automotive User Interfaces and Interactive Vehicular Applications (AutomotiveUI ’24), September 22--25, 2024, Stanford, CA, USA}
\usepackage{array}
\usepackage{caption}
\usepackage{multirow}
\usepackage{ragged2e}
\usepackage{booktabs}
\usepackage{tabularray}
\usepackage{caption}
\usepackage{subcaption}
\usepackage{algorithm2e}
\usepackage{bm}
\usepackage{amsmath}
\usepackage{enumitem}
\usepackage{hyperref}

\DeclareMathOperator*{\argmin}{arg\,min}
\DeclareMathOperator{\dis}{d}

\begin{document}
\title[Post-ride feedback to improve prosocial behavior in micromobility interactions]{Can we enhance prosocial behavior? Using post-ride feedback to improve micromobility interactions}


\author{Sidney T. Scott-Sharoni}
\email{sidney.scott.sharoni@gatech.edu}
\affiliation{%
  \institution{Georgia Institute of Technology}
  \streetaddress{School of Psychology}
  \city{Atlanta}
  \state{Georgia}
  \country{USA}
  \postcode{30332}
}
\author{Shashank Mehrotra}
\email{shashank_mehrotra@honda-ri.com}
\affiliation{%
  \institution{Honda Research Institute USA., Inc.}
  \streetaddress{70 Rio Robles}
  \city{San Jose}
  \state{California}
  \country{USA}
  \postcode{95134}
}
\author{Kevin Salubre}
\email{kevin_salubre@honda-ri.com}
\affiliation{%
  \institution{Honda Research Institute USA., Inc.}
  \streetaddress{70 Rio Robles}
  \city{San Jose}
  \state{California}
  \country{USA}
  \postcode{95134}
}
\author{Miao Song}
\email{miao_song@honda-ri.com}
\affiliation{%
  \institution{Honda Research Institute USA., Inc.}
  \streetaddress{2420 Oak Valley Drive,}
  \city{Ann Arbor}
  \state{Michigan}
  \country{USA}
  \postcode{48103}
}
\author{Teruhisa Misu}
\email{tmisu@honda-ri.com}
\affiliation{%
  \institution{Honda Research Institute USA., Inc.}
  \streetaddress{70 Rio Robles}
  \city{San Jose}
  \state{California}
  \country{USA}
  \postcode{95134}
}
\author{Kumar Akash}
\email{kakash@honda-ri.com}
\affiliation{%
  \institution{Honda Research Institute USA., Inc.}
  \streetaddress{70 Rio Robles}
  \city{San Jose}
  \state{California}
  \country{USA}
  \postcode{95134}
}
\renewcommand{\shortauthors}{Scott-Sharoni, et al.}

\begin{abstract}

Micromobility devices, such as e-scooters and delivery robots, hold promise for eco-friendly and cost-effective alternatives for future urban transportation. However, their lack of societal acceptance remains a challenge. Therefore, we must consider ways to promote prosocial behavior in micromobility interactions. We investigate how post-ride feedback can encourage the prosocial behavior of e-scooter riders while interacting with sidewalk users, including pedestrians and delivery robots. Using a web-based platform, we measure the prosocial behavior of e-scooter riders. Results found that post-ride feedback can successfully promote prosocial behavior, and objective measures indicated better gap behavior, lower speeds at interaction, and longer stopping time around other sidewalk actors. The findings of this study demonstrate the efficacy of post-ride feedback and provide a step toward designing methodologies to improve the prosocial behavior of mobility users. 
\end{abstract}



\begin{teaserfigure}
  \centering
  \includegraphics[width=.75\textwidth]{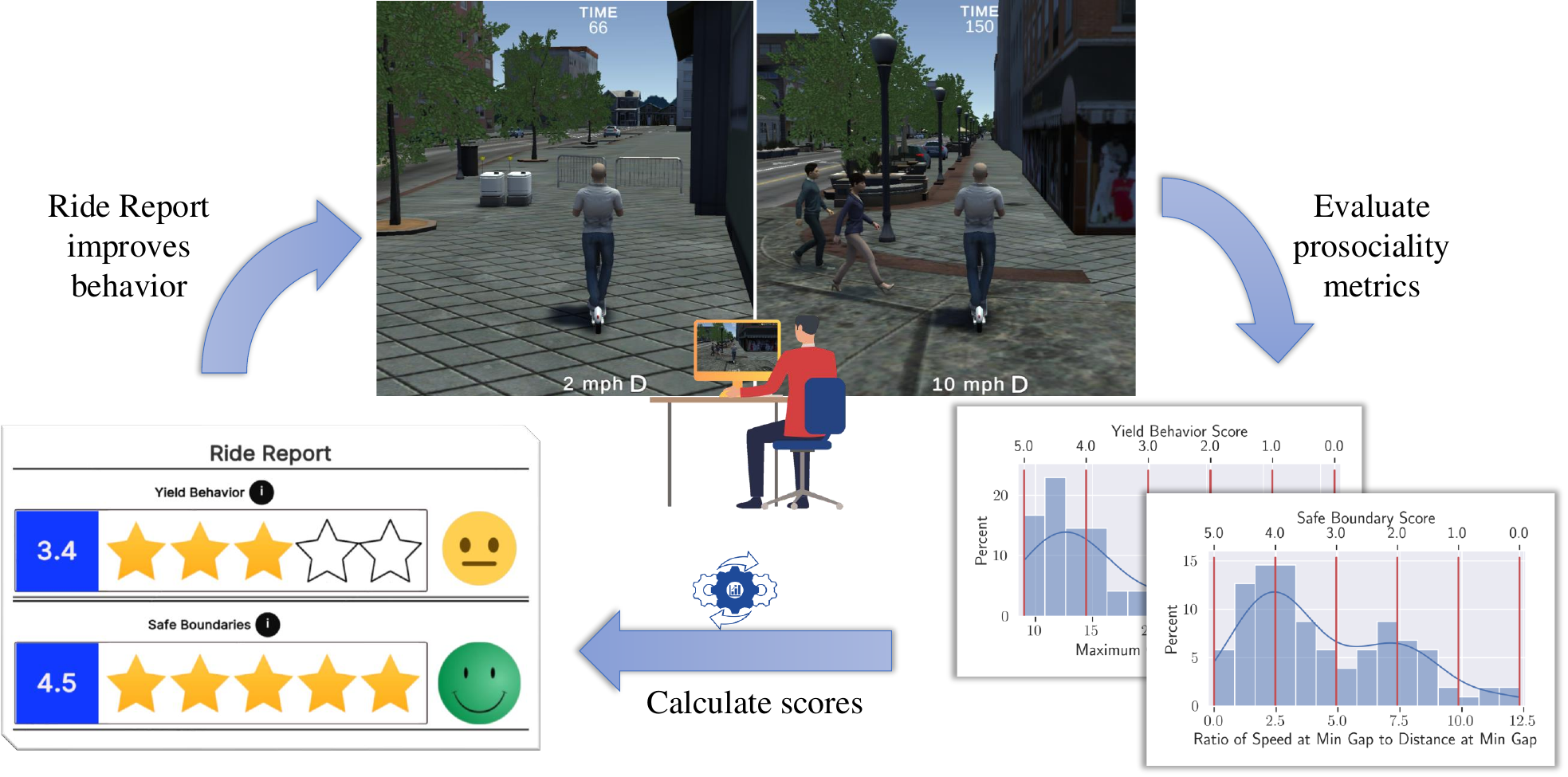}%\vspace{-8pt}
  \caption{Using post-ride feedback to improve prosocial interactions in mobility. Top left: Ego scooter yielding to automated delivery robots while maintaining a safe boundary. Top right: Ego scooter restricting the path of pedestrians. Bottom right: Scores calculation from distribution of prosociality metrics. Bottom left: Proposed prosocial behavior ride report based on user behavior.}
  \Description{Using post-ride feedback to improve prosocial interactions in mobility. Top left: Ego scooter yielding to automated delivery robots while maintaining a safe boundary. Top right: Ego scooter restricting the path of pedestrians. Bottom right: Score calculation from distribution of prosociality metrics. Bottom left: Proposed prosocial behavior ride report based on user behavior during the online study trial.}
  \label{fig:teaser}
\end{teaserfigure}

\received{17 April 2024}

\renewcommand{\shortauthors}{Scott-Sharoni, et al.}
\maketitle

\section{Introduction}

% Motivation  
% Advances in micromobility. Ban of cars in city center, more micromobility interaction. New advances in automation may reduce safety, but social acceptance is issues given human behavior
An increasing number of cities are beginning to shift their mobility solutions away from cars toward more environmentally-friendly mobility means \cite{nieuwenhuijsen2016car,niu2024beyond}. One of the most promising solutions has been micromobility, which provides flexible, sustainable, cost-effective, and on-demand transport alternatives \cite{shaheen2020sharing,mehrotra2023trust}. Micromobility has the potential to shift away from private cars significantly and help with many of the transportation-related issues that cities throughout the globe are now dealing with. With the significant rise in popularity of these micromobility modes, especially with the shared mobility model \cite{shaheen2020sharing}, there has been an increased pushback in their social acceptance. Shared electric scooters were recently banned in Paris with 90 percent support \cite{Guardian23}. Furthermore, recent research has found that users feel apprehensive about micromobility means of transportation \cite{Bretones2023, BRETONES2022}. Additionally, the rise in delivery robots, which are likely to increase traffic on sidewalks \cite{Lee2021}, may present challenges for pedestrian safety \cite{jennings2019}. %With limited space, more mobility technologies on sidewalks can complicate interactions. 

% Problem definition
Advances in driving automation technologies promise to improve sidewalk interaction safety \cite{subramanyam2023accident}. Perception and planning algorithms currently being developed for cars can potentially improve the safety of sidewalk mobility modes, such as e-scooters, delivery robots, etc. However, sidewalk micromobility modes may still be human-controlled in the near future. Therefore, promoting prosocial behavior amongst micromobility users becomes critical to ensure social acceptance of these technologies.
% Gaps and contribution
Penner et al. define prosocial behavior as ``a broad category of acts that are defined by some significant segment of society and/or one’s social group as generally beneficial to other people'' \cite{Penner2005}. Recent work has explored prosocial behavior in driving-related research, like allowing another vehicle to merge \cite{Harris2014,KAYE2022}. However, this has not been explored in a micromobility context on sidewalks where on-road traffic rules do not necessarily apply. 

In this paper, we investigate prosocial behavior in micromobility, objectively measure it, and facilitate the promotion of prosocial behavior using post-ride feedback. This allows mobility users to meaningfully self-reflect and improve prosocial behavior in future interactions. Existing research has shown that individualized feedback can improve drivers' engagement and eco-driving performances \cite{DONMEZ2008, BROUWER2015}. Extending this idea, we propose that post-ride feedback based on metrics can promote prosocial behaviors in sidewalk interactions. %To validate our method, we conduct a user study using a web-based driving simulator platform and demonstrate that the proposed post-ride feedback is successfully able to improve the prosocial behavior of micromobility users.
% We designed and conducted a two-phase user study based on a 3D user simulator with crowd workers. Based on the statistics from the first study, we designed an algorithm to measure people's prosocial behavior. The second study used the algorithm to provide post-ride feedback to another group of participants. We show that the proposed post-ride feedback is successfully able to improve the prosocial behavior of micromobility users. The overall research methodology is shown in Figure~\ref{fig_flow}.
% \begin{figure}[h]
%     \centering
%     \includegraphics[width=.9\textwidth]{Images/fig_flow.pdf}
%     \caption{Overall flow of the research methodology. }
%     \label{fig_flow}
% \end{figure}
The contributions of this paper are as follows:
% \begin{enumerate}[label={Contribution \alpha*}]
%     \item \emph{Contribution 1:} We assess the feasibility of a novel post-ride feedback focused on improving the prosocial behavior of mobility users. The feedback utilizes objective measures to generate a report of how users interacted with other road users. 
%     \item \emph{Contribution 2:} The research evaluate how prosocial behavior is influenced by environmental interactions (infrastructure or type of actors), psychological factors of the individual (personality traits), and demographic factors (age or gender). This is complemented by an analysis of self-reported subjective feedback. Through this exploration, the research validates whether the prosocial behaviors exhibited by the participants had a relationship with self-reported subjective indicators.
% \end{enumerate}
\begin{enumerate}[leftmargin=*]
    \item We demonstrate the feasibility of using post-ride feedback to improve prosocial behavior in the context of micromobility users. The feedback includes objective measures to systematically report the sidewalk interaction behaviors.  
    \item We evaluate how prosocial behavior is influenced by environmental interactions (infrastructure or type of actors). This is further complemented by an analysis of the participants' subjective responses. %Through this exploration, the research validates whether the prosocial behaviors exhibited by the participants had a relationship with self-reported subjective indicators.
\end{enumerate}

\section{Related Work}\label{sec:background}
%The following sections examine research from three overarching topics that are pertinent to our study. As our research aims to improve daily transportation and relations between human users and AI, we begin by discussing works related to human factors problems with Human-Automation Interaction (HAI) in mobility, and more specifically, micromobility. Then, we detail studies that informed our study design on prosocial behavior, as encouraging prosocial behavior can improve future mobility interactions. Finally, we include what previous researchers have found regarding post-ride feedback, including its definition and its impact on human performance.  

%\subsection{Human-Automation Interaction in Micromobility}
The advancement of novel automated vehicle (AV) technology and newer forms of shared automated vehicles (SAV) hold promise in providing benefits to people in several aspects, including daily commutes \cite{HASAN2021} \cite{Clements2017}, vehicle emissions \cite{STERN2019}, transportation affordances \cite{ALESSANDRINI2015}, and general road safety \cite{Tafidis2022}. However, the mass adoption of these technologies is unlikely to occur for at least another ten years, according to industry experts \cite{forbes2022}. In the meantime, it remains imperative for researchers to make progress on ensuring seamless interactions in these mixed autonomous environments, where emerging technology does not cause disruptions. One such transportation method is micromobilities, which are smaller, slower, and often more individual means of transportation \cite{sengul21}. Micromobility users engage in closer interactions with pedestrians than cars and may adopt different roles (i.e., pedestrian, scooter, vehicle) depending on what is most beneficial to them at each interaction with other pedestrians, sometimes at the expense of other pedestrians \cite{TUNCER2020}. This may cause discomfort to others due to potential safety risks and space-sharing conflict \cite{maiti22} \cite{GEHRKE2023} \cite{weinberg2023}. \citeauthor{Markkula2020} \cite{Markkula2020} defined space-sharing conflict as ``an observable situation from which it can be reasonably inferred that two or more road users intend to occupy the same region of space at the same time in the near future.'' We define interactions as behavioral differences between road users during a space-sharing conflict. While these definitions specify ``road users,'' we extend these concepts to the sidewalk as similar space-sharing conflicts can occur between micromobility users and pedestrians. The next section discusses the characterization of prosocial behavior in mobility. 

\subsection{Prosocial Behavior in Mobility}
Past studies on prosocial behavior typically involve games in which participants make either real or pretend decisions about how much money to give themselves and another player \cite{Thielmann2020}. While prosocial behavior has been explored in other contexts, we focus on prosocial behavior in transportation and mobility. \citeauthor{KAYE2022} \cite{KAYE2022} found that prosocial driving behavior was influenced by factors like self-esteem, efficacy, social and moral norms, and situational contexts. Social norms, such as peers telling each other to drive slowly, increased prosocial driving behaviors. \citeauthor{ward2020} \cite{ward2020} found that drivers' prosocial behavior is correlated positively with positive attitudes, normative perceptions, and perceived control. Another study concluded that prosocial driving is intentional and that research should investigate methods of increasing prosocial behaviors rather than reducing risky behaviors \cite{ward2020}. \citeauthor{Harris2014} \cite{Harris2014} created a self-report measure of prosocial and aggressive driving behaviors and found that their prosocial driving factor correlated with fewer traffic violations. They defined prosocial driving as ``a pattern of safe driving behaviors that potentially protect the well-being of passengers, other drivers, and pedestrians, and that promotes effective cooperation with others in the driving environment.'' This definition serves as the basis of our research interest in examining prosocial behavior in micromobility. To the best of our knowledge, no prior study has explored prosocial behavior in a micromobility context.

\subsection{Behavior Change through Feedback}
Research on how feedback impacts human behavior has existed for several decades, beginning steadily around the 1950s. The most fundamental research to this effect is rooted in the work by Thorndike's law of effect \cite{Thorndike_1898}, where positive feedback increases a desired behavior while negative responses can reduce it. People typically perform a desired behavior better when they are provided with knowledge of their successes and failures \cite{Preyer1959} \cite{Illgen1979}. Grounded in this established field, recent advances have demonstrated how AI-based technologies can engage young learners by emoting effectively through prosocial prompting \cite{xu2023examining}.
%While feedback-based approaches have always shown promise in behavior change, feedback does not guarantee improvement. Past meta-analysis in student learning found feedback being highly contextual on the desired outcome, type, and direction of feedback may have little-to-none or negative effects on a desired behavior \cite{Wisniewski2020}. This requires assessing learning behaviors through sensing and connected technologies, which can help systems effectively cater to the individuals' learning.  
% Feedback research using AI technologies have also demonstrated the effectiveness. For example, visualization of discussion using AI (transcription and analysis) demonstrated improved meeting learning. 
% https://link.springer.com/article/10.1007/s10639-023-12097-6
% In mobility domain, recent advancement of sensing and connected technologies have enabled various feedback methods. 
Past literature has established how prosocial behavior can be influenced through feedback. Through the utilization of the law of effect, positive emotional rewards can lead to prosocial behavior, which can further reinforce positive emotions \cite{aknin2018positive}. Behren et al., \cite{behrens2019interplay} found that immediate feedback in a group setting resulted in participants volunteering and performing cooperative behaviors in cooperative game settings. Pillitlai et al. \cite{pillutla1999social} found that participants engage in prosocial behavior in group-based settings where they receive immediate feedback on cooperative or competitive behaviors, which could be attributed to the compulsion to adhere to normative behaviors. These studies show the importance of feedback in eliciting prosocial behavior. In mobility, information about the environment can improve future driver behavior around other road actors \cite{Ma2021} \cite{FORSTER2017}. The studies on feedback design in the mobility context try to improve driving behavior by reducing hard braking events, reducing distraction, and improve takeover performance \cite{zheng2022identification, Donmez2021, DONMEZ2008, mehrotra2022human}. However,  existing research does not look to bridge the gap between post-ride feedback, micromobility, and prosocial behavior. By understanding the nuances of prosociality and reinforcing good interactions,  seamless interactions between sidewalk users can help facilitate comfortable and efficient mobility environments. This research aims to bridge these gaps through an online experimental study. %\vspace{-11pt} %Informative and individualized feedback through Human-Machine Interface (HMI) has been found to ameliorate transitions of control from the AI performing the driving function in an automated vehicle to the human user \cite{Richardson2018} \cite{Li2019} \cite{BROUWER2015}. %Another study utilized individualized feedback on HMIs to inform drivers about their eco-driving performances \cite{BROUWER2015}. 
\section{User study}\label{sec:userstudy}
To evaluate the prosocial behaviors of micromobility users, we use an online web-based driving simulator platform where the participants interact with other road users while actively controlling a micromobility. This section details the online study platform used for the evaluation, a description of the user study design and participant group, the experimental procedure, quantitative metrics for evaluating prosocial behavior, and the design of the post-ride feedback.%\vspace{-8pt} %The core elements of the user study were identical in Phase I and Phase II, as described in this paper. The only difference in the procedure involved the addition of the post-ride feedback.
\subsection{Apparatus - Online Web Study platform}
The study was conducted online using a web-based driving simulator that allowed users active longitudinal and lateral control of the e-scooter. The study trials were built in the game developing platform, Unity (v. 3.4.2) \cite{haas2014history}, using a collection of publicly purchasable 3D environmental assets amalgamated into a cohesive map of a suburban city environment. Each participant completed five trials (1 practice and 4 study trials) on the same city map but never on the same strip of sidewalk. The online format allowed the participants to complete the study through their personal desktop/laptop computers, where the study ran on a web browser.  Participants controlled a player, a person riding on an electric scooter, through the virtual city using their keyboard keys. %Participants could accelerate, decelerate, reverse, and steer left and right. 
The rider (hereafter, called \emph{ego}) appeared from a third-person camera angle (see Figure \ref{fig:teaser}). The scooter had a max speed of 10 MPH (16.1 KPH) in accordance with laws from the District of Columbia \cite{cicchino2021severity}. Participants were also advised to drive on the right side of the sidewalk when possible. To see the complete list of instructions, access the full study at \url{https://researchhost.github.io/prosocialfeedback\_review/}.%\vspace{-8pt}%\footnote{The study will be available at a permanent URL after acceptance.}
%\subsection{Experiment Design}
%For evaluating the research questions, we propose a $2\times2\times2$ mixed-design study. The between-subject factor was whether the participant received post-ride feedback (received or not received). The within-subject factors were: (1) Event type (robot vs. pedestrian) and (2) Interaction zone (crosswalk vs. no-crosswalk) as within-subject factors.
%An apriori power analysis found that a minimum of 146 participants was required to detect a significant effect with a power of 6 predictors in a linear mixed-effects regression model. The power analysis was done while aiming for a large effect size (d = 0.45). A $2\times2\times2$ mixed-design study design was considered for this study. The between-subject factor was whether they received or did not receive feedback. Two within-subject factors were the other actors (OA) type (2 levels were Pedestrian; Robot) and the traveling path of OA (2 levels were Oncoming; Crossing). %A power of 80\% and alpha=0.05 was considered. Further details on these levels and how they appear in the study trials are described later in Section \ref{sec:study_trial_descriptions}. The experiment design is illustrated in Figure~\ref{fig_ivs}.
% \begin{figure}[h]
%     \centering
%     \includegraphics[width=.8\textwidth]{Images/fig_ivs.pdf}
%     \caption{A visual depiction of the independent variables in the experiment design. In the right figure, the red arrow denotes the traveling path of OA and the blue arrow is the path of the participant controlling an e-scooter.}
%     \label{fig_ivs}
% \end{figure}
\subsection{Participants}
%An apriori power analysis found that a minimum of 146 participants was required to detect a significant effect with a power of 6 predictors in a linear mixed-effects regression model. The power analysis was done while aiming for a large effect size (d = 0.45). 
A $2\times2\times2$ mixed-design study design was considered for this study. The between-subject factor was whether they received or did not receive feedback. Two within-subject factors were the other actors (OA) type (2 levels: Pedestrian and Robot) and the traveling path of OA (2 levels: Oncoming and Crossing). A total of 304 participants completed the study via Prolific, an online study platform with more attentive responses than other online data collection tools \cite{Douglas2023}. The criteria for participation included (1) residing in the United States, (2) possessing a valid U.S. driver's license, and (3) having normal or corrected-to-normal vision. Prolific users earned \$4, which took them between 25 to 30 minutes to complete the study and not fail the attention checks. The compensation was in line with the requirements of the prolific study platform \cite{prolificMuchShould}. Several participants reported lag issues due to the limitations of their computer devices. To account for those challenges, only participants who observed the study at a median frame rate of 24 frames per second and above \cite{narang2015detecting} were considered. %A total of 208 participants met the frame rate criteria. The distribution of the demographics for the groups that received and did not receive post-ride feedback is shown in Table \ref{tab:demographic}. 
The study was approved by the Bioethics Committee in Honda R\&D (approval code: 99HM-065H)).%\vspace{-8pt}
%A power of 80\% and alpha=0.05 was considered. 
%Further details on these levels and how they appear in the study trials are described later in Section \ref{sec:study_trial_descriptions}. %The experiment design is illustrated in Figure~\ref{fig_ivs}.
% \begin{figure}[h]
%     \centering
%     \includegraphics[width=.8\textwidth]{Images/fig_ivs.pdf}
%     \caption{A visual depiction of the independent variables in the experiment design. In the right figure, the red arrow denotes the traveling path of OA and the blue arrow is the path of the participant controlling an e-scooter.}
%     \label{fig_ivs}
% \end{figure}
%The distribution of the demographics across the two groups was similar with respect to age, gender, and ethnicity reported by the participants.  
\subsection {Experimental Instructions and Study Trials}
At the beginning of the study, participants were provided with a vignette and instructions. The vignette informed the participants about the context and their task. Participants were instructed to ride the e-scooter on the sidewalk in a straight path to the opposite end of the road within three minutes. A straight path was chosen to mitigate the challenges of remembering or seeking directions. Additionally, participants could only cross streets by riding within the crosswalk road lines to control the behavior variability. A 180-second countdown timer appeared at the top of the display. At $30$ seconds, the timer would flash red, which would continue up to $-90$ seconds before the trial automatically ended. The time pressure was induced to mimic real-life situations in which people are required to make a trade-off between acting prosocially and their own needs or priorities.  Participants were also reminded to avoid collisions. None of the instructions suggested to the participants that they needed to behave prosocially or only follow behaviors based on legal restrictions. % while also preventing participants from gamifying the experiment. %To see all the instructions, we provided the study link in the section above. 
 All participants experienced a practice trial that lasted either until they reached the opposite end of the road or until the timer expired. %The session would continue for another 90 seconds after the initial 180-second timer expired. 
The practice drive did not include any other actors or events. The trial included boundary walls to ensure participants did not deviate from the path; however, in the practice trial, the boundary walls appeared from afar invisible but as a blue grid at a close distance. For all other trials, the boundary walls were invisible. We informed participants about the purpose of the wall, ensuring participants followed the path intended for the study completion. %\vspace{-8pt}%Researchers informed participants about the difference and its purpose to illustrate that participants would need to follow the intended path. 
\subsubsection{Study trial descriptions} \label{sec:study_trial_descriptions}
The study included four main trials. The study trials occurred in a suburban downtown area with light pedestrian and delivery robot traffic. No cars drove through the environment to avoid confounding participants' yielding behavior. Each study trial included four events. A different street or direction of travel in the map was used for each trial. Events were defined as interactions with a space-sharing conflict based on the definition by Markkula et al. \cite{Markkula2020}. Half of the events occurred at an intersection for each study trial. Other events occurred due to an obstacle or blocked path. All obstacles were on the right side of the sidewalk, directly in the participant's path. Participants had to maneuver around the obstacle while two other actors (OAs) moved toward the ego. For two events, OAs were a dyad of pedestrians in direct path conflict with the other, while the other two events included OAs as a dyad of delivery robots (see Figure \ref{fig:teaser}). Note that the OAs in all trials were only pedestrians or delivery robots, and the OAs moved in a dyad of their own type for the events. All OAs moved at the same speed of 5.36 KPH (3.33 MPH). While pedestrians and delivery robots, in actuality, may not move at the same speed, we wanted to control for a potential confounding effect. To prevent ordering effects, the maps were presented in varying sequences based on a balanced Latin-square design \cite{Edwards1951}. In total, four different map orders were shown to the participants.%\vspace{-12pt} 

\subsection{Procedure}
\begin{figure*}[h]
    \centering
    \includegraphics[width=.9\textwidth]{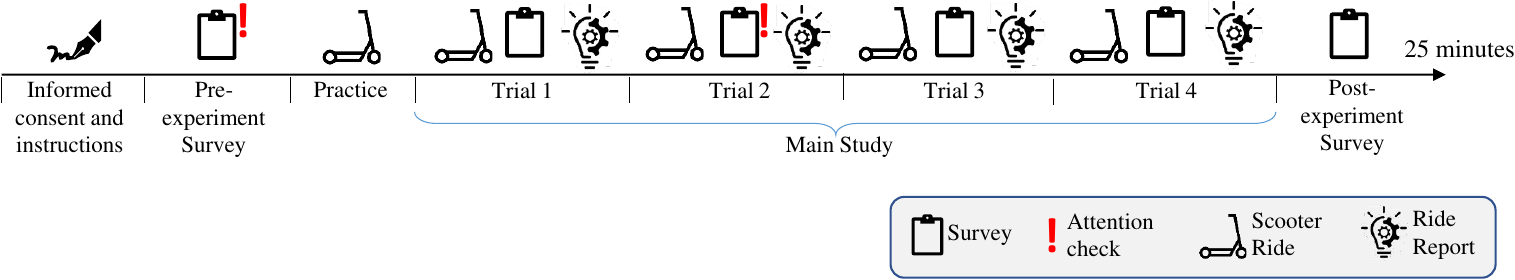}%\vspace{-8pt}
    \caption{A visual depiction of the study procedure. In the first phase of the study, the Ride Report did not occur; however, all other elements were identical.}
    \label{fig_study_design}%\vspace{-20pt}
\end{figure*}

The procedures for the two studies were nearly identical and were only differentiated by whether the participants received feedback or not (Figure \ref{fig_study_design}). Participants began the study by virtually signing the informed consent after selecting it on Prolific. Participants responded to an attestation, which asked participants to provide honest and thoughtful responses. The attestation aimed to set a cognitive mindset for the participants and prevent gamification \cite{whitson2013gaming}. Following, participants read and responded to instructions and the vignette as described in the above section.  %Participants then responded to demographic questions. These included questions regarding age, gender, and race, as well as questions regarding the mobility habits of the participants. After a short reminder of the instructions, 
A practice trial commenced to familiarize participants with the layout and controls. Participants then completed each study trial. %The participants responded to the abridged STS \cite{Ettema2011} and the perceived risk question between the trials. After these questions, individuals in the second study received post-ride feedback. 
Attention checks were included before the first trial and after the second trial. Comprehension checks appeared after the third trial and, in the second study, after the fourth trial. Post-study questionnaires recorded individual differences and ride satisfaction post-trial. The analyses of the survey responses are beyond the scope of the current article. %. After all trials were completed, participants responded to the abridged PADI and additional review questions. %Participants in the second study also received the PSSUQ and the perceived usefulness questions. 
The entire experiment for either study lasted approximately 30 minutes. 

\subsection{Quantifying Prosocial Behavior in Mobility}\label{metricsdescription}

%- what is meant by behavioral metrics
In the previous sections, we motivated the case of studying prosocial behavior in mobility. It is established that the increase in the different types of traffic participants and different micromobility would impact societal acceptance. It is, therefore, imperative to define performance measures for these behaviors. This section defines the different performance measures in detail. 

\paragraph{Minimum Gap with OA}
Past research has established how sufficient gaps between traffic participants can significantly impact perceptual decision-making when negotiating traffic situations with other actors (OA) \cite{crosato2022interaction}. Studies that have tried to understand the subjective perception of traffic participants also reported a preference for a safe gap when interacting with multiple road users \cite{mehrotra2022identifying}. With the importance of gap acceptance established, it is vital to consider how much minimum gap the ego maintains with other actors. For $\bm{x}^{ego}_t$ and $\bm{x}^{OA}_t$ be the positions of the ego and the OA at time $t$, respectively, we define the minimum gap as
$d_{min} = \min_t \hspace{2pt} \dis(\bm{x}^{ego}_t, \bm{x}^{OA}_t)$, 
where $\dis(\bm{x}_1, \bm{x}_2) = \lVert \bm{x}_1- \bm{x}_2 \rVert$ is the Euclidean distance between any positions $\bm{x}_1$ and $\bm{x}_1$.

% $$\arg \min_{d} F(d)= ( \sqrt{d}), $$  
% $$ d =(x_{OA} - x_{ego})^{2} + (y_{OA} - y_{ego})^{2} $$

\paragraph{Speed at minimum gaps}
It has been found previously that traffic participants' approach speed influences gap acceptance \cite{petzoldt2017drivers}. Perceived ego speed by other actors can greatly influence prosocial behavior and lower the participants' interaction risk \cite{Harris2014}. We define the speed at minimum gap as $v_{min} = v^{ego}_{t_{min}}$, 
where $v^{ego}_t$ is the speed of the ego at time $t$ and $t_{min} = \argmin_t \dis(\bm{x}^{ego}_t, \bm{x}^{OA}_t)$ is the time at minimum gap.

% $$\arg \min_{v} F(d)= (||v||), $$  

% $$ F(v_{mingap},d_{mingap})= (||v_{mingap}/d_{mingap}||), $$  
% $$  d_{mingap} > 0$$

\paragraph{Ego stopped time for other actors}
The metric considers the amount of time the ego spent waiting for the OAs at an event. We defined stopping when the ego's speed was less than 0.3 MPH. %Past literature has established the amount of time spent yielding to other actors can be classified as prosocial behavior \cite{Harris2014}.

\paragraph{Ego trial time}
% \subsubsection{Ego acceleration and deceleration}
The metric considers the total time a participant took to complete a trial. Although this metric does not quantify prosocial behavior, it captures the cost incurred by the participant to perform prosocial actions. Participants have to spend more trial time if they yield to OAs taking longer paths to keep more distance from OAs. 
% \subsubsection{Ego disturbance to OA}
% The metric considers the amount of disturbance caused to the OA from ego can also be considered as a reliable metric for assessing prosocial behavior. 

\section{Post-ride Feedback Design}\label{sec:feedbackdesign}
\begin{figure}[h]
 %   \centering
   \begin{subfigure}[b]{0.38\textwidth}
         \centering
         \includegraphics[width=\textwidth]
         {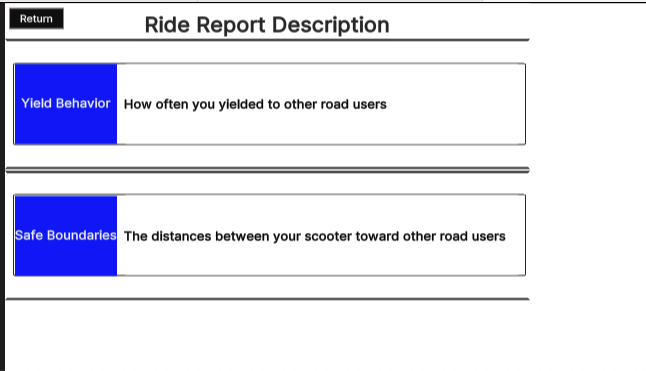}
         \caption{Description of the behaviors.}
         \label{fig: desc_info}
   \end{subfigure}
    \hspace{10pt}    
    \begin{subfigure}[b]{0.38\textwidth}
         \centering
         \includegraphics[width=\textwidth]{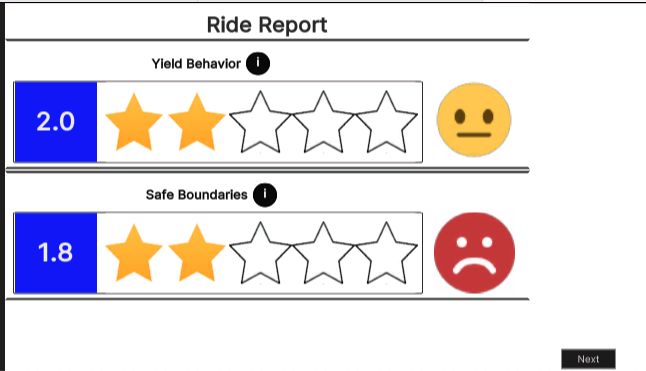}
         \caption{Example of a low-scoring report.}
         \label{fig:Low Scores on Info}
    \end{subfigure}
    \begin{subfigure}[b]{0.38\textwidth}
         \centering
         \includegraphics[width=\textwidth]{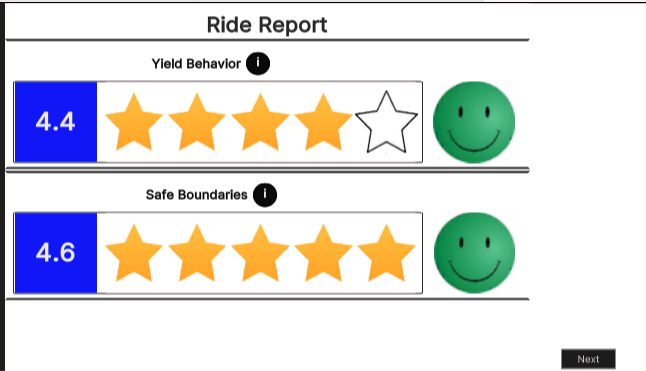}
         \caption{Example of a high-scoring report.}
         \label{fig:High Scores on Info}
    \end{subfigure}%\vspace{-8pt}
    \caption{An example of the post-ride feedback, or Ride Report, displayed to the participants.}
    \label{fig:Info Feedback}
\end{figure}

Our proposed post-ride feedback summarizes and informs participants about their prosocial behaviors. The post-ride feedback allows them to meaningfully self-reflect, thereby changing and improving their behavior in future interactions. We proposed two quantitative score-based metrics: \emph{Yield Behavior} and \emph{Safe Boundaries}; these metrics are easily comprehended and can be implemented in mobility modes that are equipped with sensors for detecting objects for collision avoidance systems. %Yielding is an example of prosociality in driving \cite{Harris2014}. However, yielding often occurs in driving as a decision to facilitate or not facilitate space. The interaction is also highly influenced by local laws. 
\emph{Yield behavior} is defined by the degree of disruption and interference placed on the others during a space-sharing conflict interaction. We calculated it based on the time OAs on the sidewalk took to cross a predefined event zone during an interaction. For example, if a participant blocked either OA's path, the OA would stop, thereby increasing their time to cross. The longer the participant blocked the path, the higher the time and, therefore, the lower the yield behavior score. Higher scores on the post-ride feedback denoted more prosociality. Participants were provided with a general definition of the yield behavior score but not to a mathematical extent (see Figure \ref{fig: desc_info}). %\emph{Yield Behavior} Another example of exhibiting prosocial behavior, specific to the micromobility context, involves the manner in which distance is kept from OAs. Passing someone at a fast speed but a large distance may be equally prosocial as passing someone at a slow speed but close distance. However, it would not be prosocial to pass someone at a fast speed but at a close distance. The degree of prosociality in interactions on a sidewalk may vary more than between drivers on the road. Therefore, we defined 
\emph{Safe boundaries} metric as the ratio of the speed of travel to the distal boundary at the minimum gap during a space-sharing conflict. Specifically, we define the safe boundary  metric as $b_{safe} = \frac{v_{min}}{d_{min}}$.

We included the word ``safe'' for the post-ride feedback design in an attempt to prevent participants from assuming that the score was solely influenced by lateral distance. The post-ride feedback, presented as the ``Ride Report'', denoted the scores numerically and visually with highlighted stars (see Figure~\ref{fig:Info Feedback}). The five stars were included so the participants could better comprehend their performance than just reporting numbers. The numerical values for the metrics were mapped to 6-point numerical metric values to scores from 0.0 to 5.0 based on the data from the no-feedback group participants. Note that the no-feedback group participants' data collection was done before the feedback group. Given an event in a route, the metric values from all the participants in the no-feedback group were first filtered for outliers based on the 1.5 IQR rule \cite{iqr}. The scores from 0.0 to 5.0 were linearly mapped from the maximum to minimum values, respectively. A sample mapping for the first event in Route 4 is shown in Figure \ref{fig:teaser}; the minimum safe boundary metric of 0.01 was mapped to a 5.0 safe boundary score,  the maximum safe boundary metric of 12.33 was mapped to a 0.0 safe boundary score, and the intermediate values were linearly mapped. Note that the lower values of the yield behavior time and safe boundary metric meant more prosocial behavior and were mapped to higher prosocial scores.
% The score for each behavior varied from 0.0 to 5.0 and was determined by XYZ. 
%- how the raw values were divided into stars  
% include that: The scoring was average of the scores based on the specific interaction events, which would then averaged with the other scores of the other events in the study trial. Score presented reflected the average score for each prosocial behavior.  

% \begin{figure}[h]
%  %   \centering
%    \begin{subfigure}[b]{0.36\textwidth}
%          \centering
%          \includegraphics[width=\textwidth]
%          {Images/fig_T4_S1_YieldBeh.pdf}
%          \caption{Mapping yield behavior time to yield behavior score.}
%          \label{fig:scores_map_yield}
%    \end{subfigure}
%     \hspace{10pt}    
%     \begin{subfigure}[b]{0.36\textwidth}
%          \centering
%     \includegraphics[width=\textwidth]{Images/fig_T4_S1_Bound.pdf}
%          \caption{Mapping safe boundary metric to safe boundary score.}
%          \label{fig:scores_map_bound}
%     \end{subfigure}
%     \caption{Sample distribution of the metrics used to calculate the post-ride feedback scores for Route 4 Event 1. The bars and the curve show the histogram and the kernel density estimate of the distribution. For a given event, based on the data distribution for a metric, the scores from 5.0 to 0.0 were linearly mapped from the minimum to maximum values of the metric, respectively.}
%     \label{fig:scores_map}
% \end{figure}

\begin{figure*}[h]
    \centering
    \includegraphics[width=.92\textwidth]{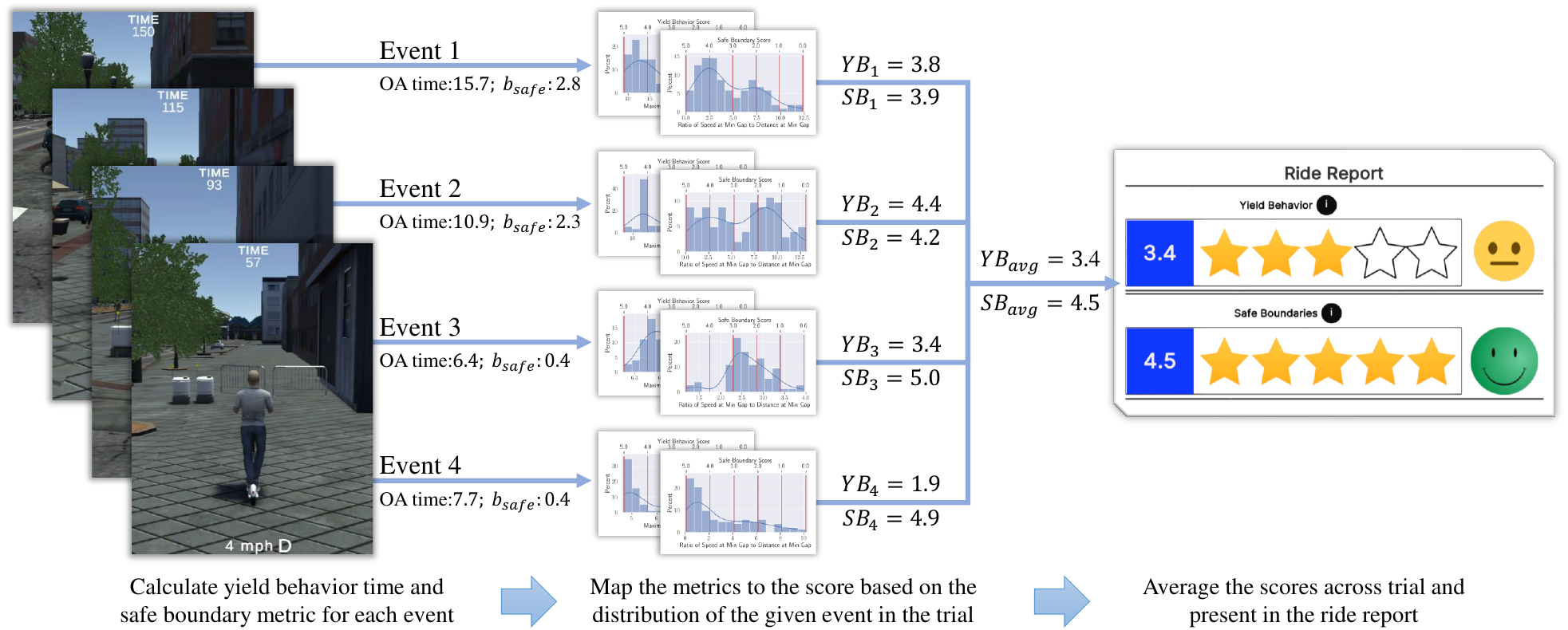}%\vspace{-8pt}
    \caption{Scores calculation for the ride report in a given trial. Sample values show the score calculation for a participant in Route 4.}
    \label{fig_scorecalc}%\vspace{-10pt}
\end{figure*}

After each trial, each user earned four scores across the four events for each behavior. The final score displayed on the Ride Report was the average value of these four scores rounded to one decimal place. Furthermore, the number of stars was based on the rounded integer values of the scores. The overall procedure to calculate the scores is shown in Figure~\ref{fig_scorecalc}.
To demonstrate whether the behavior was desirable, the Ride Report included three different possibilities for emotive responses (sad, happy, neutral) to the scoring (see Figure \ref{fig:Info Feedback}). The emotive responses differed based on the score that the participant received. Scores above 4.0 elicited a green, happy face. Scores from 2.0 to 3.9 received a yellow, neutral face. Scores below 2.0 received a red, sad face.  

\section{Findings}\label{sec:results}
We present the findings from the data analysis to evaluate the efficacy of the post-ride feedback. Firstly, we explore how the post-ride feedback influenced the scores for subsequent rides. Secondly, we detail how post-ride feedback and event-related factors in trials impacted the behavioral metrics. %Thirdly, we examine whether participants' individual differences, demographics, or subjective measures impacted the behavioral metrics. 
Finally, we report the participants' reflections on the post-ride feedback. 
\subsection{Change in Prosocial Behavior Scores across Trials} \label{subsection:feedbackperf} 
\begin{figure}[h]
    \centering
    \begin{subfigure}[b]{0.3\textwidth}
         \centering
         \includegraphics[width=\textwidth]{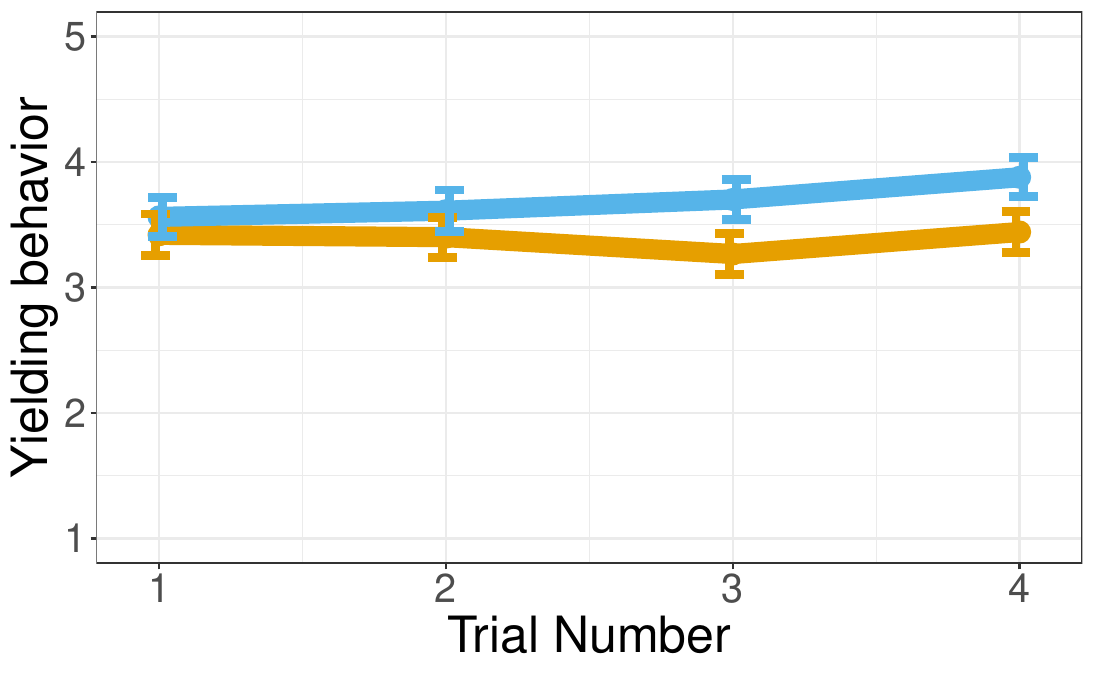}
         \caption{Yielding Behavior}
         \label{fig:YieldingBehavior}
    \end{subfigure}
    \hspace{10pt}    
    \begin{subfigure}[b]{0.3\textwidth}
         \centering
         \includegraphics[width=\textwidth]{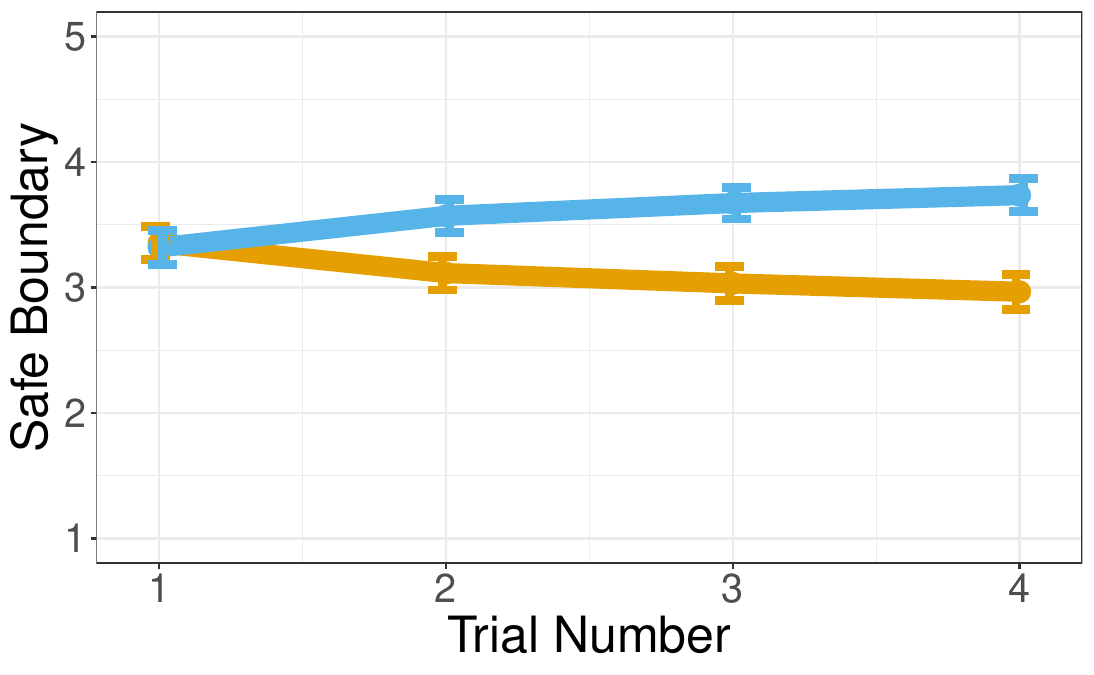}
         \caption{Safe Boundary}
         \label{fig:SafeBoundary}
    \end{subfigure}
    \\ %\vspace{3pt}
    \begin{subfigure}[b]{0.3\textwidth}
         \centering
         \includegraphics[width=\textwidth]{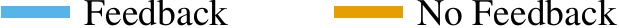}
    \end{subfigure}%\vspace{-8pt}
    \caption{Trend of scores across trials for the feedback and no-feedback group participants. }
    \label{fig:label1}%\vspace{-10pt}
\end{figure}

%- discuss the animations and illustrations that were generated 
%- the goal is to prove that the other people could do this as well and that the data is worthwhile Across the trials, participants who received feedback showed consistently higher yielding behavior and safe boundary scores than those who did not.
To compare the scores of the feedback group participants', we calculated the no-feedback group participants' scores. Given that the no-feedback group participant data was collected without the scores for the post-ride feedback. The scores were calculated post hoc. The scores were compared with the feedback group participants. We conducted four two-sample t-tests, one for each trial to compare between the two groups.  To control for an inflated familywise error rate, we used a Bonferroni Correction and only considered $p$-values as significant if they were less than 0.0125. The findings from the t-test are reported in Table \ref{Tab:ttestreport}. There was a significant difference in the yield behavior scores for Trials 3 and 4 between the two groups, with higher scores for the feedback group participants. Safe boundary scores were significantly different for Trials 2, 3, and 4. Given the participants in the feedback group only saw the report \emph{after} Trial 1, no significant differences were observed between the two groups for Trial 1. This shows that the two groups had similar yielding behavior in the beginning, which changed due to the post-ride feedback after two trials. Figure \ref{fig:YieldingBehavior} shows an increasing trend in the yield behavior scores for the participants who received the post-ride feedback. Figure \ref{fig:SafeBoundary} shows that safe boundary scores continuously increased for participants who received the post-ride feedback versus those who did not receive the feedback. %There is a net positive trend for safe boundary for the participants who received the post-ride feedback. In contrast, there was a negative trend for safe boundary scores for the no-feedback group. There was a significant difference in the scores for Trials 2, 3, and 4. Again, we see no significant difference between the two groups in Trial 1. This confirms that the two groups initially had similar safe boundary behavior, which changed due to the post-ride feedback. Findings from this analysis suggest that participants who receive the post-ride feedback are improving their scores in subsequent rides, which indicates participants actively performing to improve their scores. 

\begin{table*}[h]
\centering
\caption{Two-sample t-test results for comparing the prosociality scores of the no-feedback group participant to the feedback group participants across the trials. *** indicates statistically significant results with $p$-values $<0.001$, ns indicates not significant.}%\vspace{-10pt}
\label{Tab:ttestreport}
\resizebox{.9\linewidth}{!}{%
\begin{tabular}{lccccccrrrrrl} 
\toprule
\multirow{2}{*}{Score} & \multirow{2}{*}{Trial} & \multicolumn{2}{l}{No-Feedback group} & \multicolumn{2}{l}{Feedback group} & \multirow{2}{*}{$n1$} & \multicolumn{1}{l}{\multirow{2}{*}{$n2$}} & \multicolumn{1}{l}{\multirow{2}{*}{$t$-score}} & \multirow{2}{*}{$df$} & \multirow{2}{*}{$p$-value} & \multirow{2}{*}{$p$-adjusted} & \multicolumn{1}{r}{\multirow{2}{*}{Significance}} \\
 &  & Mean & SD & Mean & SD &  & \multicolumn{1}{l}{} & \multicolumn{1}{l}{} &  &  &  & \multicolumn{1}{r}{} \\ 
\midrule
\multirow{4}{*}{Yielding Behavior} & 1 & 3.42 & 1.72 & 3.56 & 1.65 & 412 & 420 & -1.22 & 827 & 0.222 & 0.888 & ns \\
 & 2 & 3.40 & 1.65 & 3.61 & 1.72 & 412 & 420 & -1.80 & 830 & 0.072 & 0.289 & ns \\
 & 3 & 3.27 & 1.66 & 3.70 & 1.64 & 412 & 420 & -3.8 & 829 & <0.001 & <0.001 & *** \\
 & 4 & 3.44 & 1.68 & 3.88 & 1.63 & 412 & 420 & -3.8 & 828 & <0.001 & <0.001 & *** \\ 
\midrule
\multirow{4}{*}{Safe Boundary} & 1 & 3.35 & 1.37 & 3.32 & 1.40 & 412 & 420 & 0.346 & 830 & 0.729 & 1.000 & ns \\
 & 2 & 3.11 & 1.40 & 3.57 & 1.39 & 412 & 420 & -4.73 & 829 & <0.001 & <0.001 & *** \\
 & 3 & 3.03 & 1.42 & 3.67 & 1.34 & 412 & 420 & -6.71 & 824 & <0.001 & <0.001 & *** \\
 & 4 & 2.97 & 1.43 & 3.74 & 1.38 & 412 & 420 & -7.89 & 827 & <0.001 & <0.001 & *** \\
\bottomrule
\end{tabular}
}
\end{table*}

\subsection{Evaluation of the Behavioral Metrics}
We established that the post-ride feedback influenced participant behavior over the trials. To observe the performance, behavioral metrics (defined in Section \ref{metricsdescription}) were considered. In this section, we report how the behavioral metrics were influenced by post-ride feedback and the event-related factors that the participants observed in the user study. We used linear mixed models (LMMs) using R software (version 4.1) and \textit{lme4} package to analyze the relationship between the independent variables (Feedback condition, OA type, and Traveling path of OA) and dependent variable (minimum gap, speed at minimum gap, ego stopped time for OAs, unsafe boundary, and ego trial time)~~\cite{bates2009package}. Participants and trial orders were treated as random effects. The best-fit model was selected based on the lowest AIC criterion. The best-fit model was the interaction effects in every possible combination using the compare\_performance function of the performance package in R \cite{ludecke2021performance}. Findings from the mixed effects model are shown in Table \ref{Tab:LMM_gap} and \ref{Tab:LMM_time}. The model equation
for each dependent variable (DV) is given by
\begin{align}\begin{split}
    \text{DV} \sim &\text{Feedback type} + \text{OA type } + \text{ Traveling path of OA } + \\
    & \text{Traveling path of OA : OA type } + \text{Feedback type : OA type } + \\
    & \text{Feedback type : Traveling path of OA } + \\
    & (1|\text{Participant}) + (1|\text{Trial}) + \epsilon \enspace .
\end{split}\end{align}

\begin{table*}[h]
\centering
\caption{Mixed model outputs for Minimum gap and Speed at minimum gap. Significant effects ($p < 0.05$) are marked with $^*$. \label{Tab:LMM_gap}}%\vspace{-10pt}}
\resizebox{.9\linewidth}{!}{
\begin{tabular}{lrrrrrrrr}
\toprule
\multirow{2}{*}{Independent Variable}                          & \multicolumn{4}{c}{Minimum gap}                                                 & \multicolumn{4}{c}{Speed at minimum gap}\\
                          \cmidrule{2-9}
              & \multicolumn{1}{c}{Est $\beta$} & \multicolumn{1}{c}{SE} & \multicolumn{1}{c}{t} & \multicolumn{1}{c}{p} & \multicolumn{1}{c}{Est $\beta$}         & \multicolumn{1}{c}{SE} & \multicolumn{1}{c}{t} & \multicolumn{1}{c}{p}\\
\midrule
Feedback Type (With feedback)               & $0.513$ &  $0.256$  & $2.002$ & $0.045^*$ & $-0.813$ &  $0.117$  & $-6.973$ & $<0.001^*$ \\
OA type (Robot)                                       & $0.596$  &   $0.257$ &  $2.315$ &  $0.020*$ & $-0.322$ & $0.117$ & $-2.755$ & $0.006^*$ \\
Traveling path of OA (Crossing)                 & $2.714$ &     $0.257$ &  $10.550$ &   $<0.001*$ & $0.788$  & $0.089$ & $60.701$  & $<0.001^*$ \\
Feedback Type (With feedback): OA type (Robot) & $-0.224$ & $0.296$ & $-0.755$ & $0.450\enspace$ & $0.178$ & $0.135$ & $1.321$ & $0.187\enspace$ \\
Feedback Type (With feedback): Traveling path of OA (Crossing) &  $1.100$ & $0.2961$ & $3.714$ & $<0.001^*$ & $0.513$ & $0.135$ & $3.814$ & $0.001^*$ \\
 Traveling path of OA (Crossing): OA type (Robot) & $0.984$ & $0.296$ & $3.323$ & $<0.001^*$ & $-0.024$ & $0.135$ & $-0.180$ & $0.857\enspace$ \\
\bottomrule
\end{tabular}
}
\end{table*}

\subsubsection{Behavior at gap} \label{behavior at gap}
 Based on the model results, main effects were observed for a higher minimum gap for the feedback group ($\beta$=$0.513$) at the crosswalk ($\beta$=$2.714$) and while interacting with Robots ($\beta$=$0.596$). Main effects were also observed for speed at the minimum gap for the feedback group ($\beta$=$-0.813$), and the minimum gap at the crosswalk ($\beta$=$5.427$), and when OA was a robot ($\beta$ = -$0.322$). For the minimum gap, Interaction effects were observed between Feedback received and traveling at the crosswalk ($\beta$ = $1.100$) and between traveling at the crosswalk and interacting with Robots ($\beta$ = $0.984$).  Interestingly, an interaction effect was observed for the speed at the minimum gap between participants who received the feedback and traveling at the crosswalk ($\beta$ = $0.513$).  
 
\begin{figure}[h]
    \centering
    \begin{subfigure}[b]{0.45\textwidth}
         \centering
         \includegraphics[width=\textwidth]{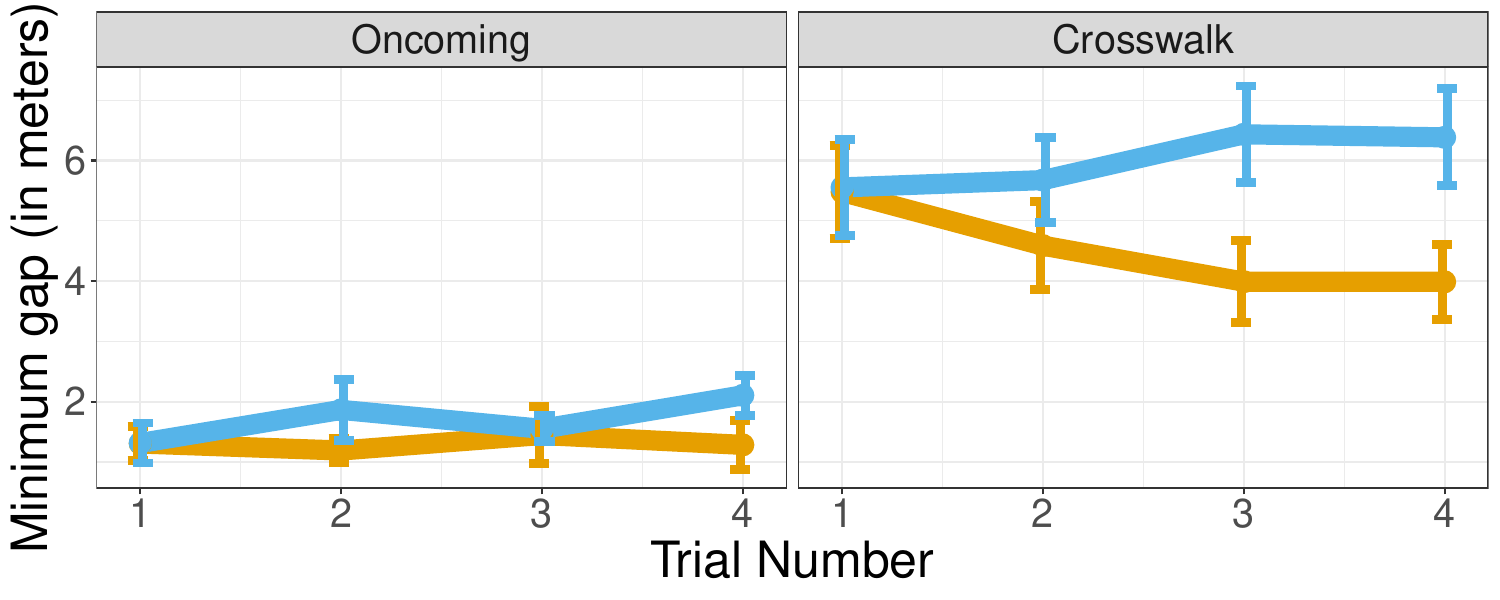}
         \caption{Variation by traveling path of OA}
         \label{fig:BM_minimumgap_OAtravel}
    \end{subfigure}
    \hspace{0.7cm}
    \begin{subfigure}[b]{0.45\textwidth}
         \centering
         \includegraphics[width=\textwidth]{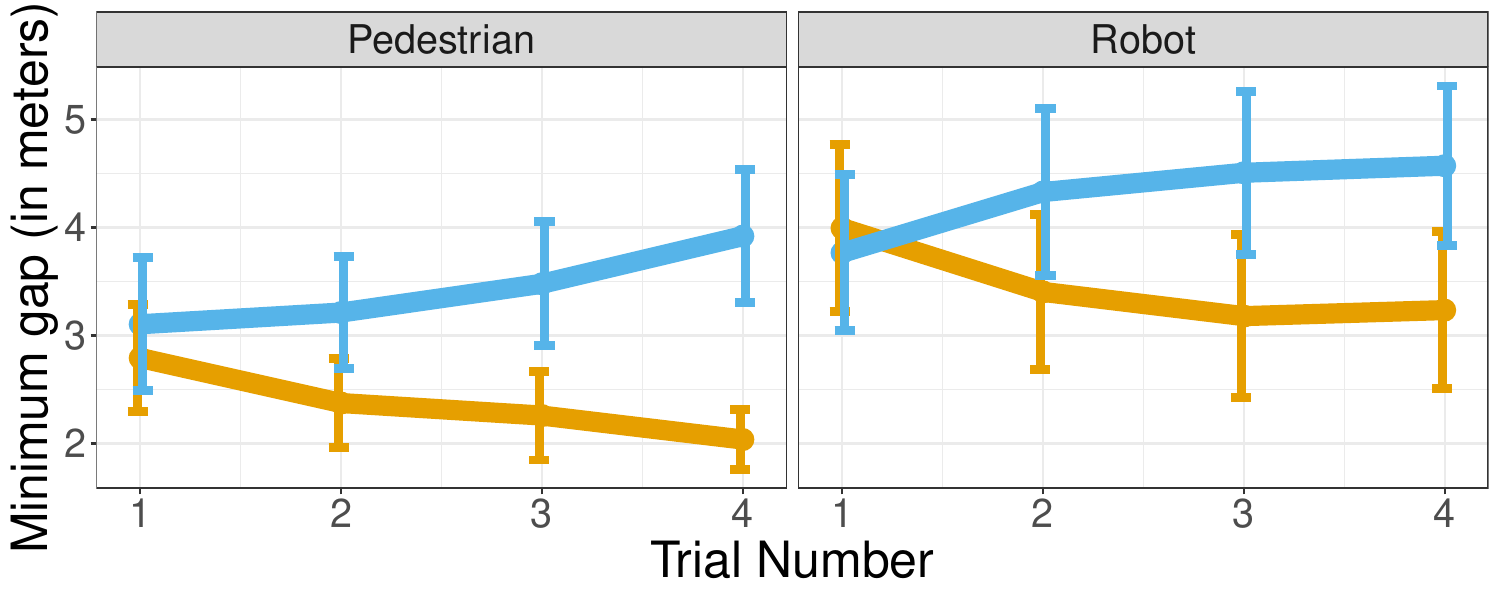}
         \caption{Variation by OA type}
         \label{fig:BM_minimumgap_OAtype}
    \end{subfigure}
    % \hspace{10pt}    
    \\ %\vspace{3pt}
    \begin{subfigure}[b]{0.3\textwidth}
         \centering
         \includegraphics[width=\textwidth]{Images/fig_legend.pdf}
    \end{subfigure}%\vspace{-8pt}
    \caption{Trends of minimum gap metric across the trials for feedback group and no-feedback group participants.}\label{fig:BM_minimumgap}%\vspace{-10pt}
\end{figure}

 Figure \ref{fig:BM_minimumgap} shows the trends for the minimum gap with OAs increased for participants who received the post-ride feedback. The trends are supported by the results from the model, which indicate the efficacy of the feedback on how participants maintained distance from other agents. The trend is prominent when the OAs are at the crosswalk. Similarly, there is a higher minimum gap with OAs for participants who received feedback. These trends are confirmed by the modeling results, which indicate that feedback influenced the minimum gap. Considering the travel path of the OAs in Figure \ref{fig:BM_minimumgap_OAtravel}, we see that the participants who received the post-ride feedback kept a higher minimum gap when encountering OAs at crosswalks. The result is expected as the crosswalks typically have larger spaces to maneuver around than block-level interactions. The minimum gap is observed to increase and maintain across subsequent trials in the crosswalk for participants who received feedback. Receiving post-ride feedback and traveling at crosswalks observed an increase in the minimum gap over successive trials for the feedback groups, as shown in Section \ref{subsection:feedbackperf}. No trends are observed for oncoming OAs. This could be attributed to a limitation in the space to negotiate, and the minimum gap may not entirely represent the minimum gap observed by the participants.%  Analyzing the speed at the minimum gap may be more meaningful in this case. 

\begin{figure}[h]
    \centering
    \begin{subfigure}[b]{0.45\textwidth}
         \centering
         \includegraphics[width=\textwidth]{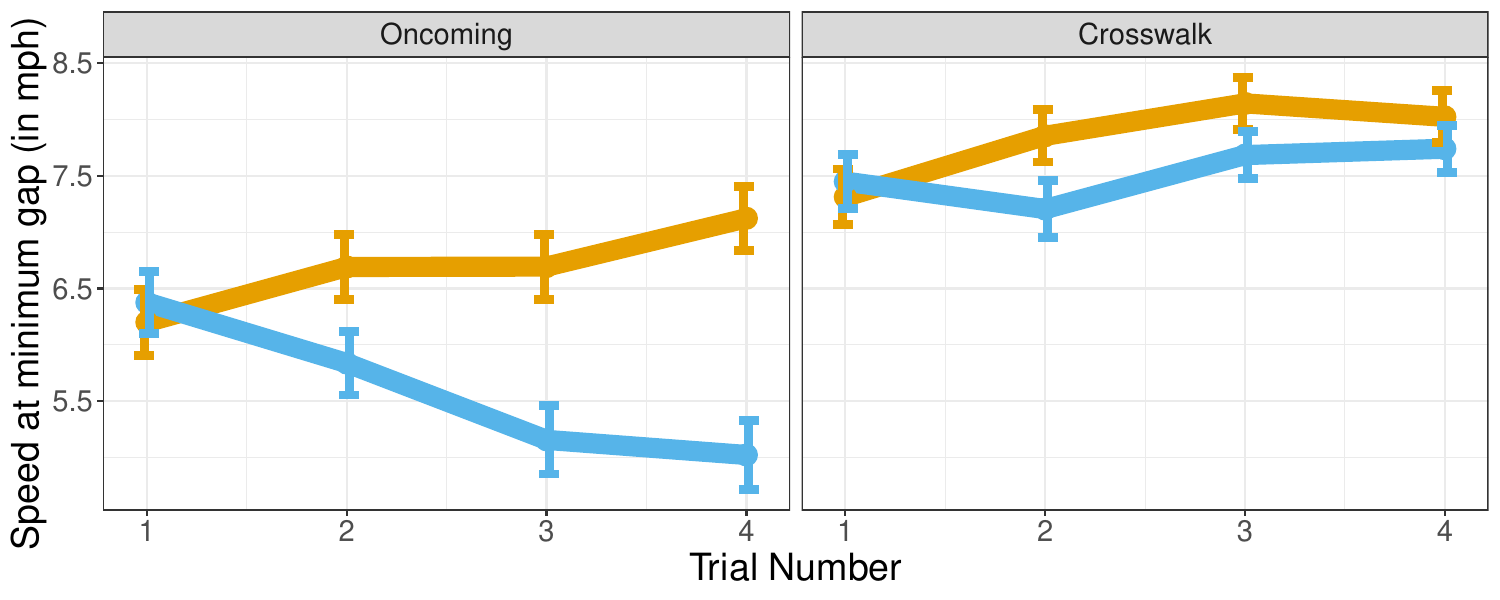}
         \caption{Variation by traveling path of OA}
         \label{fig:BM_speedgap_OAtravel}
    \end{subfigure}
    \hspace{0.7cm}
    \begin{subfigure}[b]{0.45\textwidth}
         \centering
         \includegraphics[width=\textwidth]{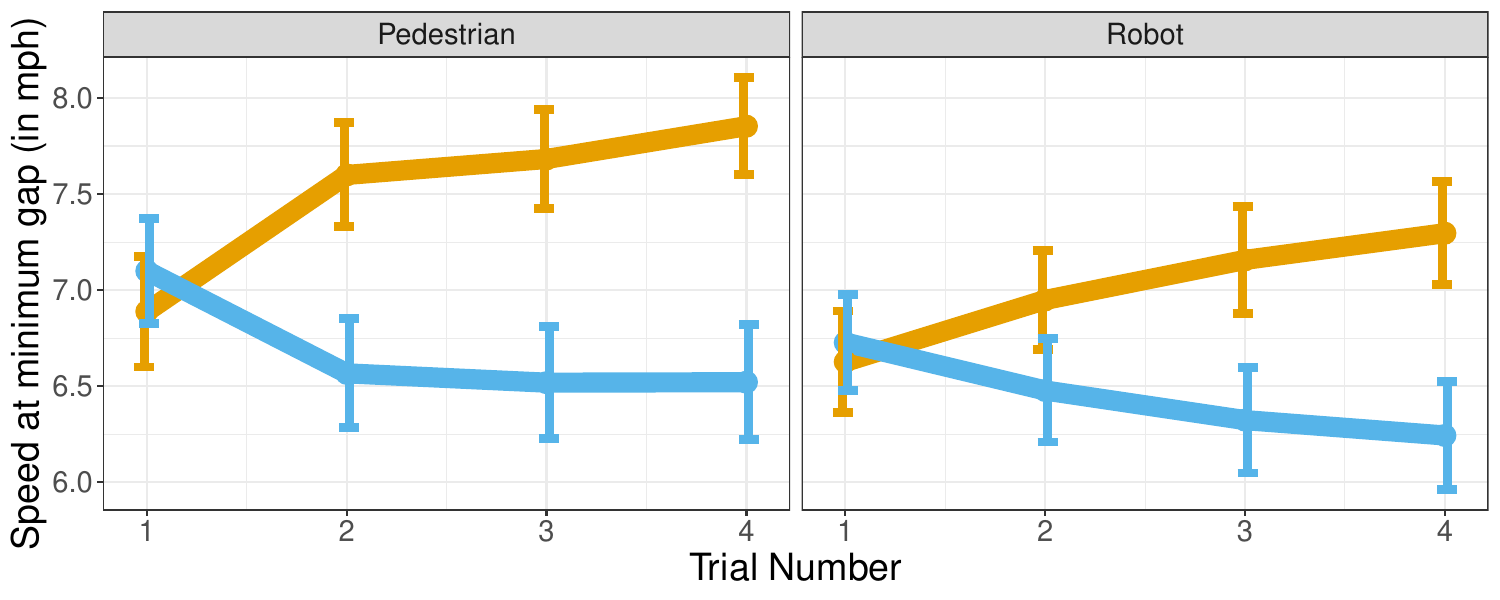}
         \caption{Variation by OA type}
         \label{fig:BM_speedgap_OAtype}
    \end{subfigure}
    % \hspace{10pt}    
    \\ %\vspace{3pt}
    \begin{subfigure}[b]{0.3\textwidth}
         \centering
         \includegraphics[width=\textwidth]{Images/fig_legend.pdf}
    \end{subfigure}%\vspace{-8pt}
    \caption{Trends of speed at minimum gap metric across the trials for feedback group and no-feedback group participants.}\label{fig:BM_speedgap}%\vspace{-15pt}
\end{figure}

% Figure \ref{fig:BM_minimumgap_OAtype} shows how participants observed a higher gap with OA type (pedestrians vs. robots) when they received the post-ride feedback. There are lower gaps with OAs when participants did not receive the post-ride feedback. We also observe that the range of values for the gap with robots is higher for robots as compared to pedestrians. This could be attributed to participants anticipating the behavior of pedestrians better than the robots. Future studies should consider whether the participants were satisfied with the trajectory of robots, although examining the perceptions of different path-planning algorithms for robots is beyond the scope of the current research. 
For speed at minimum gap, the model results also aligned with the trends observed in Figure~\ref{fig:BM_speedgap}. The trends confirm that the speed at the minimum gap decreased for the participants who received the post-ride feedback. Figure \ref{fig:BM_speedgap_OAtravel} shows different speeds at the minimum gap for different travel paths of the OAs. Participants kept a lower speed minimum gap when encountering oncoming OAs and received post-ride feedback. However, the speed at crosswalks remained higher when the travel path of OAs was crosswalk. This was further confirmed as an interaction effect was observed between receiving feedback and traveling at crosswalk. Additionally, there is a decrease in the speed with OAs, as shown in Figure \ref{fig:BM_speedgap_OAtype}. We observe slower speeds around the robots as compared to pedestrians. Irrespective of the individual interactions, the findings from the analysis confirm that the post-ride feedback does result in a lowering of the speed around OAs and lower speed with oncoming OAs, which can be indicative of better-yielding behavior whenever the opportunity to yield presents itself to the participants.

\begin{table*}[h]
\centering
\caption{Mixed model outputs for Ego stopped time for OAs and Ego trial time. Significant effects ($p < 0.05$) are marked with $^*$. \label{Tab:LMM_time}}%\vspace{-10pt}}
\resizebox{.9\linewidth}{!}{
\begin{tabular}{lrrrrrrrr}
\toprule
\multirow{2}{*}{Independent Variable}             & \multicolumn{4}{c}{Ego stopped time for OAs}  & \multicolumn{4}{c}{Ego trial time}\\                          \cmidrule{2-9}
            & \multicolumn{1}{c}{Est $\beta$} & \multicolumn{1}{c}{SE} & \multicolumn{1}{c}{t} & \multicolumn{1}{c}{p} & \multicolumn{1}{c}{Est $\beta$}         & \multicolumn{1}{c}{SE} & \multicolumn{1}{c}{t} & \multicolumn{1}{c}{p}\\
\midrule
Feedback Type (With feedback)               & $0.312$ & $0.107$ & $2.925$ & $0.004^*$ & $7.942$ & $1.628$ & $4.879$ & $<0.001^*$ \\
OA type (Robot)                                       & $0.219$ & $0.107$ & $2.047$ & $0.041^*$ & $<0.001$ & $1.633$ & $0.000$ & $1.000\enspace$ \\
Traveling path of OA (Crossing)                 & $-0.333$ & $0.107$ & $-3.110$ & $0.002^*$ & $<0.001$ & $1.633$ & $0.000$ & $1.000\enspace$ \\
Feedback Type (With feedback): OA type (Robot) & $0.072$ & $0.123$ & $0.585$ & $0.558\enspace$ & $<0.001$ & $1.879$ & $0.000$ & $1.000\enspace$ \\
Feedback Type (With feedback): Traveling path of OA (Crossing) & $-0.305$ & $0.123$ & $-2.478$ & $0.013^*$ & $<0.001$ & $1.879$ & $0.000$ & $1.000\enspace$ \\
 Traveling path of OA (Crossing): OA type (Robot) & $-0.147$ & $0.123$ & $-1.193$ & $0.233\enspace$ & $<0.001$ & $1.879$ & $0.000$ & $1.000\enspace$ \\
\bottomrule
\end{tabular}
}
\end{table*}

\subsubsection{Ego stopped time}
\begin{figure}[h]
    \begin{subfigure}[b]{0.45\textwidth}
         \centering
         \includegraphics[width=\textwidth]{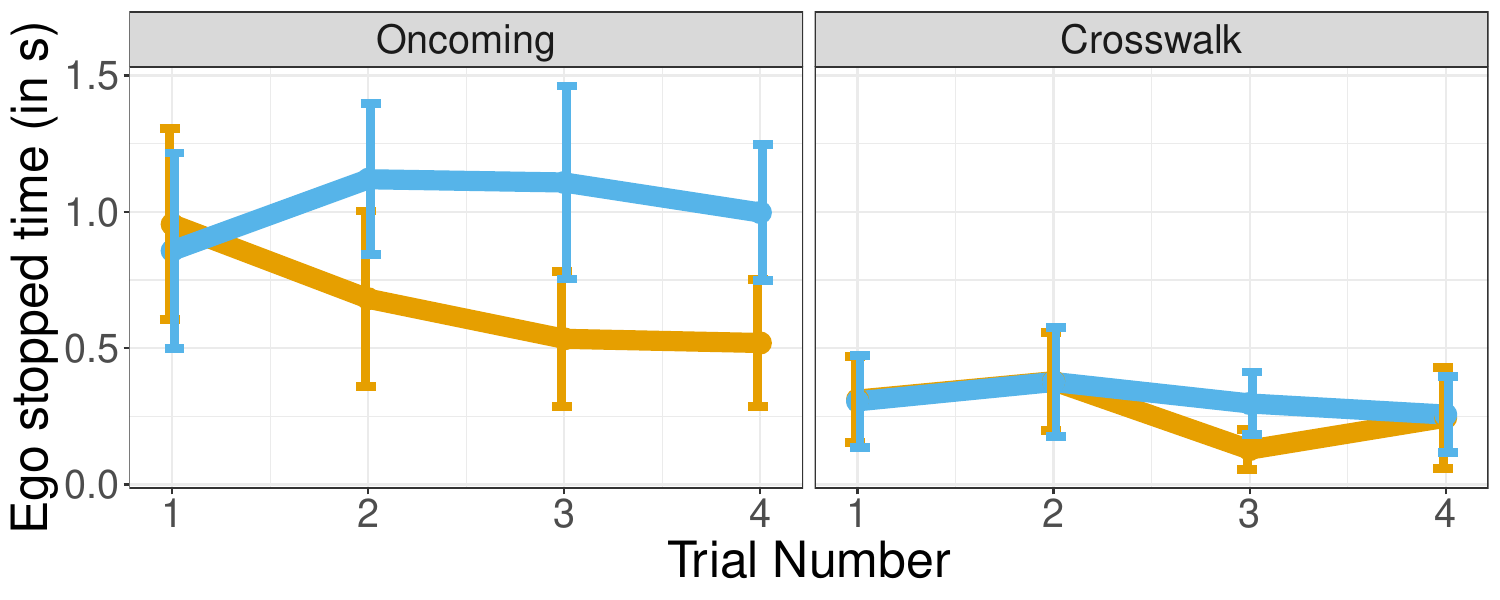}
         \caption{Variation by traveling path of OA}
         \label{fig:BM_egostoppedtimescene}
    \end{subfigure}
    \hspace{0.7cm}
    \begin{subfigure}[b]{0.45\textwidth}
         \centering
         \includegraphics[width=\textwidth]{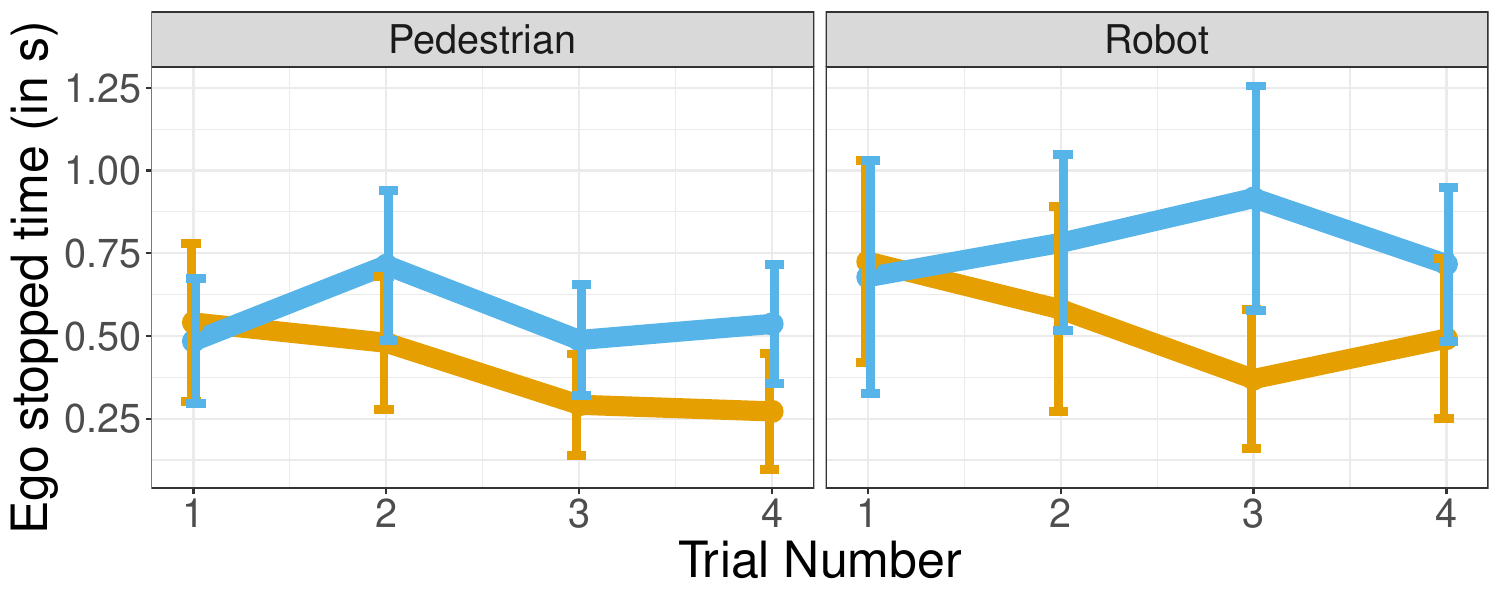}
         \caption{Variation by OA type}
         \label{fig:BM_egostoppedtimeagent}
    \end{subfigure}
    \\ %\vspace{3pt}
    \begin{subfigure}[b]{0.3\textwidth}
         \centering
         \includegraphics[width=\textwidth]{Images/fig_legend.pdf}
    \end{subfigure}%\vspace{-8pt}
    \caption{Trends of ego stopped time metric across the trials for feedback group and no-feedback group participants.}\label{fig:BM_egostoppedtime}%\vspace{-12pt}
\end{figure}

 Figure \ref{fig:BM_egostoppedtime} shows that ego stopped time increased for the participants who received the post-ride feedback. Figure \ref{fig:BM_egostoppedtimescene} shows the distribution of the ego time spent stopping for different travel paths of OAs. Participants generally stopped for a shorter time at the crosswalk than for oncoming OAs. For oncoming OAs, feedback group participants stopped longer than the no-feedback group. This effect can be attributed to post-ride feedback causing participants to proactively stop for OAs, which aligns with the findings in Section \ref{behavior at gap}. For OAs, the ego stopped time is slightly higher for Robots, which could be attributed to participants cautiously assessing the trajectory of the delivery robots. In contrast, they would better understand the expected behaviors of the pedestrians. Stopping behavior is similar for pedestrians and robots. The trends can be seen in Figure \ref{fig:BM_egostoppedtimeagent}.

\subsubsection{Ego trial time}
\begin{figure}[h]
    \begin{subfigure}[b]{0.45\textwidth}
         \centering
         \includegraphics[width=\textwidth]{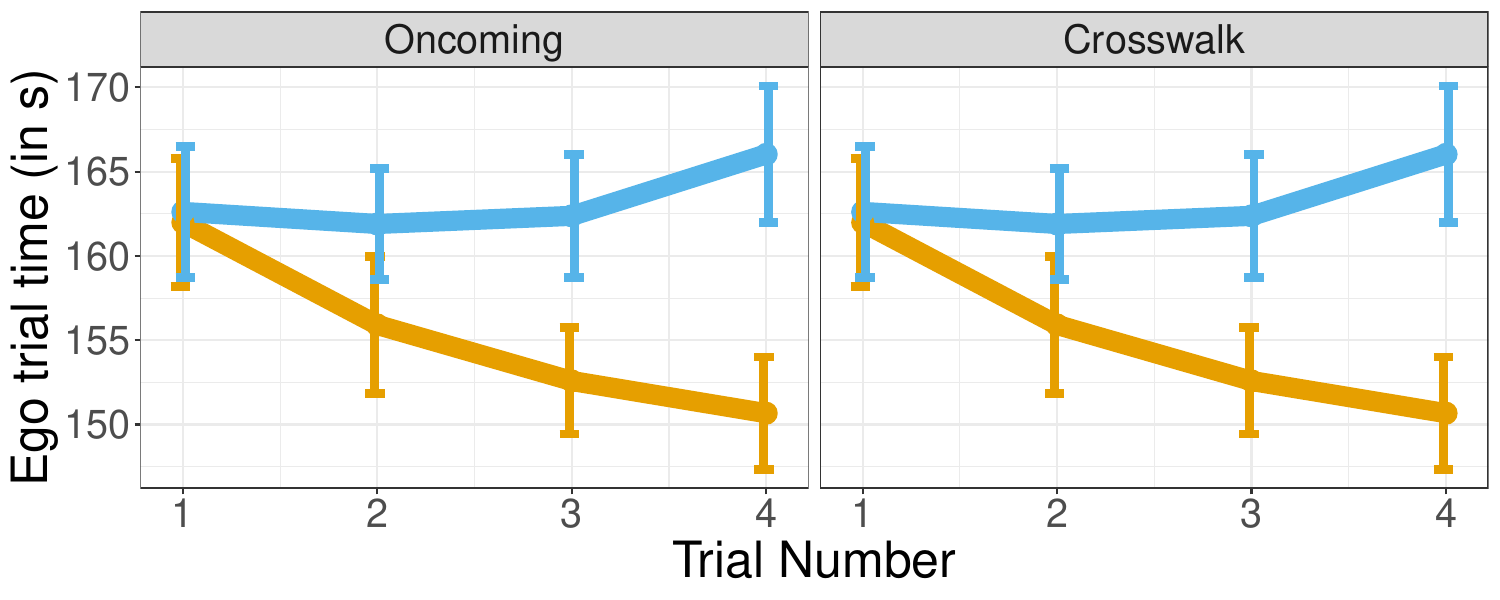}
         \caption{Variation by traveling path of OA}
         \label{fig:BM_trialcompletiontimescene}
    \end{subfigure}
    \hspace{0.7cm}
    \begin{subfigure}[b]{0.45\textwidth}
         \centering
         \includegraphics[width=\textwidth]{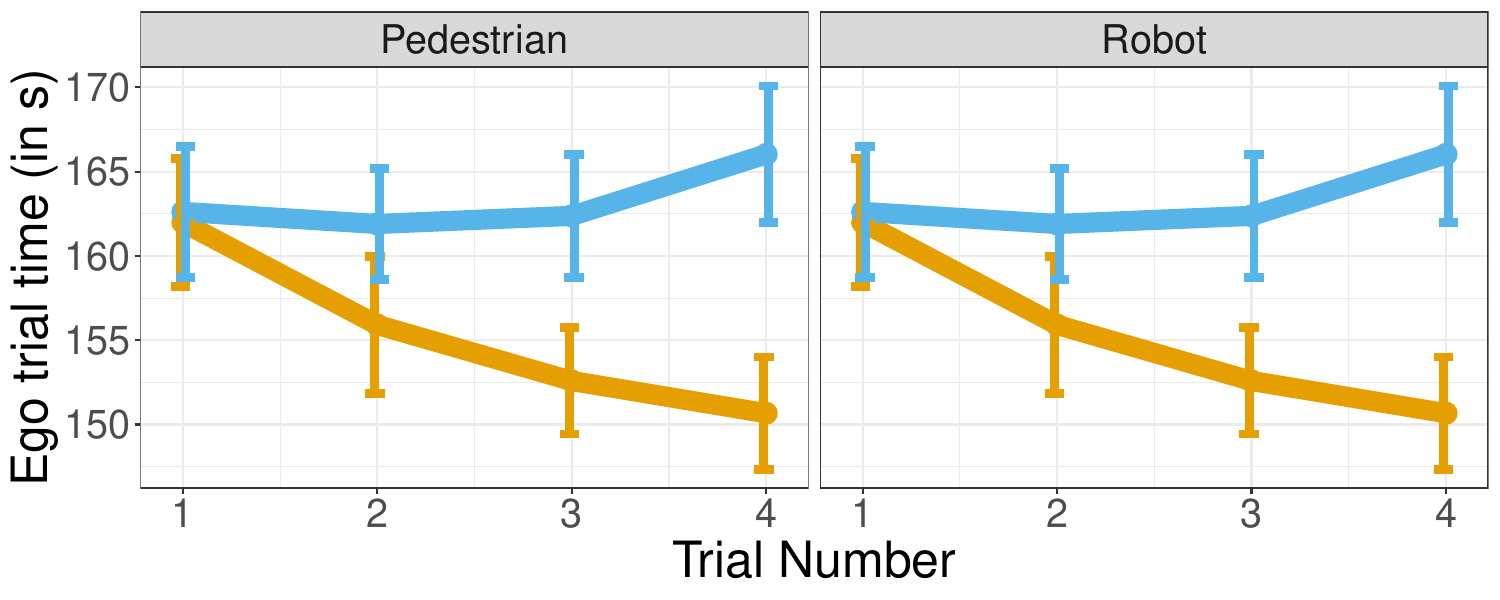}
         \caption{Variation by OA type}
         \label{fig:BM_trialcompletiontimeagent}
    \end{subfigure}
    \\ %\vspace{3pt}
    \begin{subfigure}[b]{0.3\textwidth}
         \centering
         \includegraphics[width=\textwidth]{Images/fig_legend.pdf}
    \end{subfigure}%\vspace{-8pt}
    \caption{Trends of ego trial completion time metric across the trials for feedback group and no-feedback group participants.}\label{fig:BM_egotrialcompletion}%\vspace{-15pt}
\end{figure}

From Figure~\ref{fig:BM_egotrialcompletion}, we see that the ego trial completion time increased for the participants who received post-ride feedback. % across all OA types and traveling paths of OAs.
 From the estimates in Table~\ref{Tab:LMM_time} and Figure~\ref{fig:BM_egotrialcompletion}, it is worth noting that post-ride feedback led to an increase in trial completion time. %Participants who did not receive the post-ride feedback completed the trials faster. 
 %This indicates that post-ride feedback aided participants in prioritizing prosocial behavior compared to those who did not receive the post-ride feedback. 
 While trial completion may not be representative of prosocial behavior, it is indicative of costs associated with performing prosocial behaviors. Note that the participant pool of prolific workers typically optimizes time efficiency in task completion~\cite{gadiraju2019crowd}. While participants were neither incentivized for timely completion nor penalized for delay in completion, participants did forego time voluntarily. %It is interesting because the participants are prioritizing improving their prosociality above timeliness. 
 We also expect this phenomenon in real-world situations, where people may prioritize prosociality in dyadic interactions over trip completion when aware of their behaviors. Given that people do not incur significant loss due to acting prosocially, we believe that such feedback can increase prosocial behavior in people.

\subsection{Impact of the post-ride feedback on the trials}
It was evident that changes in scores occurred throughout the trials, which is also supported by the subjective responses from the participants. Based on a frequency analysis, most participants ($77$/$105$) reported actively trying to change their behavior to increase their scores. However, a small number of participants ($18$/$105$) reported no influence on their behavior, which could indicate that the post-ride feedback did not necessarily influence the way they rode. The rest of the participants' responses ($10$/$105$) were not conclusive.  Despite this, an increase in yielding behaviors and safe boundaries were the most cited responses on how participants changed their scores, highlighting the impact of the post-ride feedback of the ride report. As participant P14 noted, 
\emph{``I tried to keep my previous rating in mind for the next trial each time.  I did my best to improve as much as possible.  When I saw an obstacle, I tried slowing to around 5 MPH before approaching it.  This gave me more time to react if necessary.''}. 
Another participant P29 remarked,
\emph{``I simply tried to follow the rules of the road more carefully on subsequent ride tests. For example, I tried to yield to pedestrians more often to get a better ride score. I would stop in the middle of a crosswalk to prevent getting too close to other robots or pedestrians on the other side of the sidewalk path.''}%\vspace{-8pt}

\paragraph{Sentiments about the scores}
Subjective opinions of the scores were categorized into positive and negative sentiments. Most participants ($42$/$105$) had a positive opinion toward the feedback. However, a significant population ($39$/$105$) had mixed feelings about the scores they received. Participants expressed skepticism about the scores, with opinions citing the scores' inaccuracy, harshness, and unfairness. Additionally, participants believed the scores they received did not match their expectations based on their rides. Overall, it was evident that a lack of transparency in the scoring system caused skepticism toward the report, which may have resulted in some participants not caring about the feedback. As Participant P92 explained, 
\emph{``I felt that the scores I received were a little critical as I did not collide with any pedestrians and yet still made it to the end of the street in time. I prioritized reaching the end of the street within the time limit as I came very close to not making it in time in a few trials, with only a few seconds to spare. I would have liked to be more considerate while traveling on the sidewalk, but the time limit was short enough to prevent me from doing things like coming to a complete stop at times.''} Additionally, participant P11 stated, \emph{``I wasn't sure if they were random or actually based on my performance. I wanted to complete each ride as quickly as possible without bumping into others, so I tried to avoid others rather than yield to them narrowly. I didn't care about the effect on my score.''}. The participants' opinions provided important perspectives on post-ride feedback. They expressed interest in understanding the cause of the performance scores. The feedback provides detailed insights on how the post-ride feedback can potentially help participants reflect on their reports and could potentially aid in further improving the efficacy of the post-ride feedback. %Participant P38 commented,
\section{Discussion} \label{sec:discussion}
% 1. Summary of the results - 1 paragraph
% 2. Theoretical/General discussion required 
This study assessed the feasibility of post-ride feedback improving the prosocial behavior of mobility users. The feedback utilized objective measures and reported how users interacted with other road users. A total of 208 participants' data was considered in the online study, where 105 participants received the post-ride feedback, and 103 participants received no feedback. The study evaluated how prosocial behavior is influenced by mobility interactions. % (infrastructure or type of actors), individual (prosociality), and demographic factors (age and gender). The self-reported subjective feedback was analyzed to consider how participants perceived the feedback and how it impacted their behaviors. 
%\vspace{-6pt}
%Findings from the study con 
\subsection{Influence of feedback on prosocial behavior}
The analysis found that participants who received post-ride feedback maintained larger gaps and increased stop time while interacting with OAs. They also lowered their speed at oncoming traffic paths. Overall, participants also increased the overall trial time. The improvement in prosocial behavior using informative feedback suggests that feedback can help improve mobility behaviors, which is supported by previous findings \cite{eckoldt2016gentleman,wang2016likes,knobel2013become}. Eckoldt et al. \cite{eckoldt2016gentleman} suggested that designing for prosocial driving behaviors enables better communication and smoother interactions with other road users. 
Another finding from our study demonstrated that informative feedback resulted in the participants lowering their speed when the available space was low (oncoming traffic) and increasing the gap when the available space was higher (crosswalk). This aligns with the work done by Harris et al. \cite{Harris2014}, where they considered maintaining sufficient space, lower speeds close to other road actors, and stopping for others when required as suitable proxy behaviors for assessing prosocial behavior in mobility environments.%\vspace{-11pt} 
\subsection{Participant opinions on the Feedback report}
Participants expressed engagement with the post-ride feedback and proactively looked to engage in prosocial behavior based on the reports. Most of the participants engaged with the report, indicating post-ride feedback efficacy. While most participants expressed positive opinions on the post-ride feedback, there were concerns about a lack of transparency on the scoring system and a lack of recommendations on improving their behavior. In the free-form response texts, several participants provided suggestions to enhance the efficacy of the post-ride feedback by (1) clarifying the scoring scheme to enhance transparency and (2) including supplemental information to help identify ways to improve pro-social behavior. These findings provide invaluable insights into this feasibility study and an opportunity to improve the future study with a user-oriented post-ride feedback system.

The study format presented the opportunity for gamification; however, gamification is a strong tool for increasing desired behaviors \cite{KRATH2021}, and similar methods have been used to reduce unsafe driving behaviors \cite{forbes_23}. %The notion that positive feedback increases a behavior while negative responses can reduce a behavior stems back to Thorndike's Law of Effect \cite{Thorndike_1898}.
The negative responses from the informative feedback caused participants to amend and decrease non-prosocial behaviors such as short gap distances and blocking the path of OAs. To amend the prosocial scores after receiving feedback, participants likely required increased attention and effort, as suggested by the increase in trial time. The findings from the study are supported by Feedback Intervention Theory (FIT), which can guide attention to behavior to meet the standards. This can explain the increased trial time in individuals who received post-trial feedback. Regardless of whether the primary motivator was an internal desire to behave prosocially or to receive a higher score, the feedback motivated participants to behave differently. However, understanding the true motivations behind prosocial behavior may be challenging. %\vspace{-6pt}%Batson \cite{BATSON1987} posited that prosocial behaviors are motivated by intrinsic egoistic desires or external altruistic desires to help others. Whether feedback improved prosocial behavior due to altruism or wanting to achieve a certain score is pertinent, but understanding how to produce the desired behavior could help understand future mobility interactions.

\subsection{Limitations and future work}\label{sec:limitation}
While this research evaluates the efficacy of objective measures for prosocial behavior and the feasibility of post-ride feedback, there are some limitations of this research. The online study provides a low-fidelity platform for evaluating behaviors in a gaming-like experience. We plan to conduct follow-up VR-based evaluations to mimic close-to-real-world behaviors in the near future. In-person study could further benefit from the contextualization of the perceived rewards in time-demanding situations and whether participants would engage in prosocial behavior if they experienced the temporal demands of commuting in a real-world setting. Additionally, the evaluation performance can be confounded by prior experience participating in simulator environments like studies or playing video games, which could be accounted for in future studies. % experienced in an in-person setting and provide further explanations of how the feedback would influence behaviors %as the perception of risks, consequences, and prosocial behavior may not be possible to simulate.
%Additionally, this feasibility study had limited interactive OAs to control for confounding effects. However, other scenarios with different numbers of OAs are needed to generalize the results further. Future work looks to replicate the study in a lab study and validate the measures in a naturalistic setting. These replications would help provide further validation of these concepts. Furthermore, given the online participant pool, we could not control for experiences with micromobility in the participants. Given that the experiences with micromobility can impact participants' behaviors, future work should look at its effect on prosocial behavior. Another limitation of the study is the minimal nature of the post-ride feedback. Participants expressed their opinions about the lack of transparency and recommendations of the post-ride feedback. Our future work will consider incorporating the participants' opinions to design more interactive feedback. 
%\vspace{-3pt}
\section{Conclusion} \label{sec:conclusion}
%- keep this shorter and spend more time on results with a lot of graphs 
%The recent emergence of novel mobility systems like e-scooters offers opportunities to tackle global urban transportation issues and poses new challenges. 
In this work, we proposed a post-ride feedback-based methodology to promote prosocial behavior in micromobility users. An online Unity-based interactive user study was designed where the ego rider interacted with pedestrians and robots, and the performance was evaluated through objective measures that helped quantify prosocial behavior. %Additionally, subjective measures around participants' satisfaction with travel and prosocial driving behavior (PADI) were considered. 
Participants who received the post-ride feedback provided insights toward the future improvement of the feedback design. The results found that the post-ride feedback successfully improved prosocial behavior.  %Factors like age, gender, and PADI were found to influence prosocial behavior. 
The findings from this study provide a step toward designing effective post-ride feedback that facilitates safer interactions and societal acceptance of new mobility technologies. 
% 1.1.1 Contribution 1. The main contribution of the study to extant literature and future product design by assessing how post-ride feedback on mobility behaviors impacts users’ prosocial behaviors on an e-scooter. 
% Additionally, by assessing subjective indicators of prosocial driving, personality, and demographics, researchers and designers may develop systems that promote prosocial behaviors in mobility and transportation.
% 1.1.2 Contribution 2. The study also provides a demonstration of how to conduct mobility research studies in an online platform. 
% Several online studies involve only subjective data (CITE). We provide an example of how to collect behavioral data online.
%  The study included several controls that can assist other researchers in studying concepts in a convenient, affordable manner.
%  We performed the study partially as a proof of concept in that researchers can collect human subjects’ data related to transportation and mobility virtually using the Unity environment.

% Understanding methods to engender and support prosocial behavior is necessary to alleviate stresses when interacting with more advanced AI systems. This research provided an initial step into the investigation of improving prosocial behaviors in micromobility.  
\begin{acks}
We would like to thank Hiu Chun lo, Research Engineer at Honda Research Institute, who was instrumental in providing support for the design of the user study platform. %We also thank the researchers at [Anonymized Organization] for reviewing our manuscript and providing feedback.
\end{acks}
\bibliographystyle{ACM-Reference-Format}
\bibliography{main_references}


\begin{thebibliography}{60}

%%% the Web address.

\ifx \showCODEN    \undefined \def \showCODEN     #1{\unskip}     \fi
\ifx \showDOI      \undefined \def \showDOI       #1{#1}\fi
\ifx \showISBNx    \undefined \def \showISBNx     #1{\unskip}     \fi
\ifx \showISBNxiii \undefined \def \showISBNxiii  #1{\unskip}     \fi
\ifx \showISSN     \undefined \def \showISSN      #1{\unskip}     \fi
\ifx \showLCCN     \undefined \def \showLCCN      #1{\unskip}     \fi
\ifx \shownote     \undefined \def \shownote      #1{#1}          \fi
\ifx \showarticletitle \undefined \def \showarticletitle #1{#1}   \fi
\ifx \showURL      \undefined \def \showURL       {\relax}        \fi
% invisible to TeX
\providecommand\bibfield[2]{#2}
\providecommand\bibinfo[2]{#2}
\providecommand\natexlab[1]{#1}
\providecommand\showeprint[2][]{arXiv:#2}

\bibitem[pro({[n.\,d.]})]%
        {prolificMuchShould}
 \bibinfo{year}{[n.\,d.]}\natexlab{}.
\newblock \bibinfo{title}{{H}ow much should you pay research participants? | {P}rolific --- prolific.com}.
\newblock \bibinfo{howpublished}{\url{https://www.prolific.com/resources/how-much-should-you-pay-research-participants}}.
\newblock
\newblock
\shownote{[Accessed 24-06-2024]}.


\bibitem[iqr(2008)]%
        {iqr}
 \bibinfo{year}{2008}\natexlab{}.
\newblock \bibinfo{booktitle}{\emph{Interquartile Range}}.
\newblock \bibinfo{publisher}{Springer New York}, \bibinfo{address}{New York, NY}, \bibinfo{pages}{266--267}.
\newblock
\showISBNx{978-0-387-32833-1}
\urldef\tempurl%
\url{https://doi.org/10.1007/978-0-387-32833-1_200}
\showDOI{\tempurl}


\bibitem[Aknin et~al\mbox{.}(2018)]%
        {aknin2018positive}
\bibfield{author}{\bibinfo{person}{Lara~B Aknin}, \bibinfo{person}{Julia~W Van~de Vondervoort}, {and} \bibinfo{person}{J~Kiley Hamlin}.} \bibinfo{year}{2018}\natexlab{}.
\newblock \showarticletitle{Positive feelings reward and promote prosocial behavior}.
\newblock \bibinfo{journal}{\emph{Current opinion in psychology}}  \bibinfo{volume}{20} (\bibinfo{year}{2018}), \bibinfo{pages}{55--59}.
\newblock


\bibitem[Alessandrini et~al\mbox{.}(2015)]%
        {ALESSANDRINI2015}
\bibfield{author}{\bibinfo{person}{Adriano Alessandrini}, \bibinfo{person}{Andrea Campagna}, \bibinfo{person}{Paolo~Delle Site}, \bibinfo{person}{Francesco Filippi}, {and} \bibinfo{person}{Luca Persia}.} \bibinfo{year}{2015}\natexlab{}.
\newblock \showarticletitle{Automated Vehicles and the Rethinking of Mobility and Cities}.
\newblock \bibinfo{journal}{\emph{Transportation Research Procedia}}  \bibinfo{volume}{5} (\bibinfo{year}{2015}), \bibinfo{pages}{145--160}.
\newblock
\showISSN{2352-1465}
\urldef\tempurl%
\url{https://doi.org/10.1016/j.trpro.2015.01.002}
\showDOI{\tempurl}
\newblock
\shownote{SIDT Scientific Seminar 2013}.


\bibitem[Bates et~al\mbox{.}(2009)]%
        {bates2009package}
\bibfield{author}{\bibinfo{person}{Douglas Bates}, \bibinfo{person}{Martin Maechler}, \bibinfo{person}{Ben Bolker}, \bibinfo{person}{Steven Walker}, \bibinfo{person}{Rune Haubo~Bojesen Christensen}, \bibinfo{person}{Henrik Singmann}, \bibinfo{person}{Bin Dai}, \bibinfo{person}{Fabian Scheipl}, \bibinfo{person}{Gabor Grothendieck}, \bibinfo{person}{Peter Green}, {et~al\mbox{.}}} \bibinfo{year}{2009}\natexlab{}.
\newblock \showarticletitle{Package ‘lme4’}.
\newblock \bibinfo{journal}{\emph{URL http://lme4. r-forge. r-project. org}} (\bibinfo{year}{2009}).
\newblock


\bibitem[Behrens and Kret(2019)]%
        {behrens2019interplay}
\bibfield{author}{\bibinfo{person}{Friederike Behrens} {and} \bibinfo{person}{Mariska~E Kret}.} \bibinfo{year}{2019}\natexlab{}.
\newblock \showarticletitle{The interplay between face-to-face contact and feedback on cooperation during real-life interactions}.
\newblock \bibinfo{journal}{\emph{Journal of Nonverbal Behavior}} \bibinfo{volume}{43}, \bibinfo{number}{4} (\bibinfo{year}{2019}), \bibinfo{pages}{513--528}.
\newblock


\bibitem[Bretones and Marquet(2022)]%
        {BRETONES2022}
\bibfield{author}{\bibinfo{person}{Alexandra Bretones} {and} \bibinfo{person}{Oriol Marquet}.} \bibinfo{year}{2022}\natexlab{}.
\newblock \showarticletitle{Sociopsychological factors associated with the adoption and usage of electric micromobility. A literature review}.
\newblock \bibinfo{journal}{\emph{Transport Policy}}  \bibinfo{volume}{127} (\bibinfo{year}{2022}), \bibinfo{pages}{230--249}.
\newblock
\showISSN{0967-070X}
\urldef\tempurl%
\url{https://doi.org/10.1016/j.tranpol.2022.09.008}
\showDOI{\tempurl}


\bibitem[Bretones et~al\mbox{.}(2023)]%
        {Bretones2023}
\bibfield{author}{\bibinfo{person}{Alexandra Bretones}, \bibinfo{person}{Oriol Marquet}, \bibinfo{person}{Carolyn Daher}, \bibinfo{person}{Laura Hidalgo}, \bibinfo{person}{Mark Nieuwenhuijsen}, \bibinfo{person}{Carme Miralles‑Guasch}, \bibinfo{person}{}, {and} \bibinfo{person}{Natalie Mueller}.} \bibinfo{year}{2023}\natexlab{}.
\newblock \showarticletitle{Public Health‑Led Insights on Electric Micro‑mobility Adoption and Use: A Scoping Review}.
\newblock \bibinfo{journal}{\emph{Journal of Urban Health}}  \bibinfo{volume}{100} (\bibinfo{year}{2023}), \bibinfo{pages}{612–626}.
\newblock
\urldef\tempurl%
\url{https://doi.org/10.1007/s11524-023-00731-0}
\showDOI{\tempurl}


\bibitem[Brouwer et~al\mbox{.}(2015)]%
        {BROUWER2015}
\bibfield{author}{\bibinfo{person}{R.F.T. Brouwer}, \bibinfo{person}{A. Stuiver}, \bibinfo{person}{T. Hof}, \bibinfo{person}{L. Kroon}, \bibinfo{person}{J. Pauwelussen}, {and} \bibinfo{person}{B. Holleman}.} \bibinfo{year}{2015}\natexlab{}.
\newblock \showarticletitle{Personalised feedback and eco-driving: An explorative study}.
\newblock \bibinfo{journal}{\emph{Transportation Research Part C: Emerging Technologies}}  \bibinfo{volume}{58} (\bibinfo{year}{2015}), \bibinfo{pages}{760--771}.
\newblock
\showISSN{0968-090X}
\urldef\tempurl%
\url{https://doi.org/10.1016/j.trc.2015.04.027}
\showDOI{\tempurl}
\newblock
\shownote{Technologies to support green driving}.


\bibitem[Cicchino et~al\mbox{.}(2021)]%
        {cicchino2021severity}
\bibfield{author}{\bibinfo{person}{Jessica~B Cicchino}, \bibinfo{person}{Paige~E Kulie}, {and} \bibinfo{person}{Melissa~L McCarthy}.} \bibinfo{year}{2021}\natexlab{}.
\newblock \showarticletitle{Severity of e-scooter rider injuries associated with trip characteristics}.
\newblock \bibinfo{journal}{\emph{Journal of safety research}}  \bibinfo{volume}{76} (\bibinfo{year}{2021}), \bibinfo{pages}{256--261}.
\newblock


\bibitem[Clements and Kockelman(2017)]%
        {Clements2017}
\bibfield{author}{\bibinfo{person}{Lewis~M. Clements} {and} \bibinfo{person}{Kara~M. Kockelman}.} \bibinfo{year}{2017}\natexlab{}.
\newblock \showarticletitle{Economic Effects of Automated Vehicles}.
\newblock \bibinfo{journal}{\emph{Transportation Research Record}} \bibinfo{volume}{2606}, \bibinfo{number}{1} (\bibinfo{year}{2017}), \bibinfo{pages}{106--114}.
\newblock
\urldef\tempurl%
\url{https://doi.org/10.3141/2606-14}
\showDOI{\tempurl}


\bibitem[Crosato et~al\mbox{.}(2022)]%
        {crosato2022interaction}
\bibfield{author}{\bibinfo{person}{Luca Crosato}, \bibinfo{person}{Hubert~PH Shum}, \bibinfo{person}{Edmond~SL Ho}, {and} \bibinfo{person}{Chongfeng Wei}.} \bibinfo{year}{2022}\natexlab{}.
\newblock \showarticletitle{Interaction-aware decision-making for automated vehicles using social value orientation}.
\newblock \bibinfo{journal}{\emph{IEEE Transactions on Intelligent Vehicles}} \bibinfo{volume}{8}, \bibinfo{number}{2} (\bibinfo{year}{2022}), \bibinfo{pages}{1339--1349}.
\newblock


\bibitem[Donmez et~al\mbox{.}(2008)]%
        {DONMEZ2008}
\bibfield{author}{\bibinfo{person}{Birsen Donmez}, \bibinfo{person}{Linda~Ng Boyle}, {and} \bibinfo{person}{John~D. Lee}.} \bibinfo{year}{2008}\natexlab{}.
\newblock \showarticletitle{Mitigating driver distraction with retrospective and concurrent feedback}.
\newblock \bibinfo{journal}{\emph{Accident Analysis \& Prevention}} \bibinfo{volume}{40}, \bibinfo{number}{2} (\bibinfo{year}{2008}), \bibinfo{pages}{776--786}.
\newblock
\showISSN{0001-4575}
\urldef\tempurl%
\url{https://doi.org/10.1016/j.aap.2007.09.023}
\showDOI{\tempurl}


\bibitem[Donmez et~al\mbox{.}(2021)]%
        {Donmez2021}
\bibfield{author}{\bibinfo{person}{Birsen Donmez}, \bibinfo{person}{Maryam Merrikhpour}, {and} \bibinfo{person}{Mehdi~Hoseinzadeh Nooshabadi}.} \bibinfo{year}{2021}\natexlab{}.
\newblock \showarticletitle{Mitigating Teen Driver Distraction: In-Vehicle Feedback Based on Peer Social Norms}.
\newblock \bibinfo{journal}{\emph{Human Factors}} \bibinfo{volume}{63}, \bibinfo{number}{3} (\bibinfo{year}{2021}), \bibinfo{pages}{503--518}.
\newblock
\urldef\tempurl%
\url{https://doi.org/10.1177/0018720819891285}
\showDOI{\tempurl}
\newblock
\shownote{PMID: 31869571}.


\bibitem[Douglas et~al\mbox{.}(2023)]%
        {Douglas2023}
\bibfield{author}{\bibinfo{person}{Benjamin~D. Douglas}, \bibinfo{person}{Patrick~J. Ewell}, {and} \bibinfo{person}{Markus Brauer}.} \bibinfo{year}{2023}\natexlab{}.
\newblock \showarticletitle{Data quality in online human-subjects research: Comparisons between MTurk, Prolific, CloudResearch, Qualtrics, and SONA}.
\newblock \bibinfo{journal}{\emph{PLOS ONE}} \bibinfo{volume}{18}, \bibinfo{number}{3} (\bibinfo{date}{03} \bibinfo{year}{2023}), \bibinfo{pages}{1--17}.
\newblock
\urldef\tempurl%
\url{https://doi.org/10.1371/journal.pone.0279720}
\showDOI{\tempurl}


\bibitem[Eckoldt et~al\mbox{.}(2016)]%
        {eckoldt2016gentleman}
\bibfield{author}{\bibinfo{person}{Kai Eckoldt}, \bibinfo{person}{Marc Hassenzahl}, \bibinfo{person}{Matthias Laschke}, \bibinfo{person}{Thies Schneider}, \bibinfo{person}{Josef Schumann}, {and} \bibinfo{person}{Stefan K{\"o}nsgen}.} \bibinfo{year}{2016}\natexlab{}.
\newblock \showarticletitle{The Gentleman. A prosocial assistance system to promote considerate driving}. In \bibinfo{booktitle}{\emph{Proceedings on the 10th Conference on Design and Emotion}}. \bibinfo{pages}{307--314}.
\newblock


\bibitem[Edwards(1951)]%
        {Edwards1951}
\bibfield{author}{\bibinfo{person}{Allen Edwards}.} \bibinfo{year}{1951}\natexlab{}.
\newblock \showarticletitle{Balanced latin-square designs in psychological research}.
\newblock \bibinfo{journal}{\emph{The American Journal of Psychology}}  \bibinfo{volume}{64} (\bibinfo{year}{1951}), \bibinfo{pages}{598--603}.
\newblock
Issue 4.
\urldef\tempurl%
\url{https://doi.org/10.2307/1418200}
\showDOI{\tempurl}


\bibitem[Forster et~al\mbox{.}(2017)]%
        {FORSTER2017}
\bibfield{author}{\bibinfo{person}{Yannick Forster}, \bibinfo{person}{Frederik Naujoks}, \bibinfo{person}{Alexandra Neukum}, {and} \bibinfo{person}{Lynn Huestegge}.} \bibinfo{year}{2017}\natexlab{}.
\newblock \showarticletitle{Driver compliance to take-over requests with different auditory outputs in conditional automation}.
\newblock \bibinfo{journal}{\emph{Accident Analysis \& Prevention}}  \bibinfo{volume}{109} (\bibinfo{year}{2017}), \bibinfo{pages}{18--28}.
\newblock
\showISSN{0001-4575}
\urldef\tempurl%
\url{https://doi.org/10.1016/j.aap.2017.09.019}
\showDOI{\tempurl}


\bibitem[Gadiraju et~al\mbox{.}(2019)]%
        {gadiraju2019crowd}
\bibfield{author}{\bibinfo{person}{Ujwal Gadiraju}, \bibinfo{person}{Gianluca Demartini}, \bibinfo{person}{Ricardo Kawase}, {and} \bibinfo{person}{Stefan Dietze}.} \bibinfo{year}{2019}\natexlab{}.
\newblock \showarticletitle{Crowd anatomy beyond the good and bad: Behavioral traces for crowd worker modeling and pre-selection}.
\newblock \bibinfo{journal}{\emph{Computer Supported Cooperative Work (CSCW)}}  \bibinfo{volume}{28} (\bibinfo{year}{2019}), \bibinfo{pages}{815--841}.
\newblock


\bibitem[Gehrke et~al\mbox{.}(2023)]%
        {GEHRKE2023}
\bibfield{author}{\bibinfo{person}{Steven~R. Gehrke}, \bibinfo{person}{Christopher~D. Phair}, \bibinfo{person}{Brendan~J. Russo}, {and} \bibinfo{person}{Edward~J. Smaglik}.} \bibinfo{year}{2023}\natexlab{}.
\newblock \showarticletitle{Observed sidewalk autonomous delivery robot interactions with pedestrians and bicyclists}.
\newblock \bibinfo{journal}{\emph{Transportation Research Interdisciplinary Perspectives}}  \bibinfo{volume}{18} (\bibinfo{year}{2023}), \bibinfo{pages}{100789}.
\newblock
\showISSN{2590-1982}
\urldef\tempurl%
\url{https://doi.org/10.1016/j.trip.2023.100789}
\showDOI{\tempurl}


\bibitem[Giuffrida(4 02)]%
        {Guardian23}
\bibfield{author}{\bibinfo{person}{Angela Giuffrida}.} \bibinfo{year}{2023-04-02}\natexlab{}.
\newblock \showarticletitle{Parisians vote to ban rental e-scooters from French capital by huge margin}.
\newblock \bibinfo{journal}{\emph{The Guardian}} (\bibinfo{year}{2023-04-02}).
\newblock
\urldef\tempurl%
\url{https://www.theguardian.com/world/2023/apr/02/parisians-vote-on-banning-e-scooters-from-french-capital}
\showURL{%
\tempurl}


\bibitem[Haas(2014)]%
        {haas2014history}
\bibfield{author}{\bibinfo{person}{John~K Haas}.} \bibinfo{year}{2014}\natexlab{}.
\newblock \showarticletitle{A history of the unity game engine}.
\newblock  (\bibinfo{year}{2014}).
\newblock


\bibitem[Harris et~al\mbox{.}(2014)]%
        {Harris2014}
\bibfield{author}{\bibinfo{person}{Paul~B. Harris}, \bibinfo{person}{John~M. Houston}, \bibinfo{person}{Jose~A. Vazquez}, \bibinfo{person}{Janan~A. Smither}, \bibinfo{person}{Amanda Harms}, \bibinfo{person}{Jeffrey~A. Dahlke}, {and} \bibinfo{person}{Daniel~A. Sachau}.} \bibinfo{year}{2014}\natexlab{}.
\newblock \showarticletitle{The Prosocial and Aggressive Driving Inventory (PADI): A self-report measure of safe and unsafe driving behaviors}.
\newblock \bibinfo{journal}{\emph{Accident Analysis \& Prevention}}  \bibinfo{volume}{72} (\bibinfo{year}{2014}), \bibinfo{pages}{1--8}.
\newblock
\showISSN{0001-4575}
\urldef\tempurl%
\url{https://doi.org/10.1016/j.aap.2014.05.023}
\showDOI{\tempurl}


\bibitem[Hasan and {Van Hentenryck}(2021)]%
        {HASAN2021}
\bibfield{author}{\bibinfo{person}{Mohd.~Hafiz Hasan} {and} \bibinfo{person}{Pascal {Van Hentenryck}}.} \bibinfo{year}{2021}\natexlab{}.
\newblock \showarticletitle{The benefits of autonomous vehicles for community-based trip sharing}.
\newblock \bibinfo{journal}{\emph{Transportation Research Part C: Emerging Technologies}}  \bibinfo{volume}{124} (\bibinfo{year}{2021}), \bibinfo{pages}{102929}.
\newblock
\showISSN{0968-090X}
\urldef\tempurl%
\url{https://doi.org/10.1016/j.trc.2020.102929}
\showDOI{\tempurl}


\bibitem[Ilgen et~al\mbox{.}(1979)]%
        {Illgen1979}
\bibfield{author}{\bibinfo{person}{Daniel Ilgen}, \bibinfo{person}{Cynthia Fisher}, {and} \bibinfo{person}{M. Susan}.} \bibinfo{year}{1979}\natexlab{}.
\newblock \showarticletitle{Consequences of individual feedback on behavior in organizations}.
\newblock \bibinfo{journal}{\emph{Journal of Applied Psychology}}  \bibinfo{volume}{64} (\bibinfo{date}{08} \bibinfo{year}{1979}), \bibinfo{pages}{349--371}.
\newblock
\urldef\tempurl%
\url{https://doi.org/10.1037/0021-9010.64.4.349}
\showDOI{\tempurl}


\bibitem[Jennings and Figliozzi(2019)]%
        {jennings2019}
\bibfield{author}{\bibinfo{person}{Dylan Jennings} {and} \bibinfo{person}{Miguel Figliozzi}.} \bibinfo{year}{2019}\natexlab{}.
\newblock \showarticletitle{Study of Sidewalk Autonomous Delivery Robots and Their Potential Impacts on Freight Efficiency and Travel}.
\newblock \bibinfo{journal}{\emph{Transportation Research Record}}  \bibinfo{volume}{2673} (\bibinfo{year}{2019}), \bibinfo{pages}{317--326}.
\newblock
Issue 6.
\urldef\tempurl%
\url{https://doi.org/10.1177/0361198119849398}
\showDOI{\tempurl}


\bibitem[Kaye et~al\mbox{.}(2022)]%
        {KAYE2022}
\bibfield{author}{\bibinfo{person}{Sherrie-Anne Kaye}, \bibinfo{person}{David Rodwell}, \bibinfo{person}{Natalie Watson-Brown}, \bibinfo{person}{Chae Rose}, {and} \bibinfo{person}{Lisa Buckley}.} \bibinfo{year}{2022}\natexlab{}.
\newblock \showarticletitle{Road users’ engagement in prosocial and altruistic behaviors: A systematic review}.
\newblock \bibinfo{journal}{\emph{Journal of Safety Research}}  \bibinfo{volume}{82} (\bibinfo{year}{2022}), \bibinfo{pages}{342--351}.
\newblock
\showISSN{0022-4375}
\urldef\tempurl%
\url{https://doi.org/10.1016/j.jsr.2022.06.010}
\showDOI{\tempurl}


\bibitem[Knobel et~al\mbox{.}(2013)]%
        {knobel2013become}
\bibfield{author}{\bibinfo{person}{Martin Knobel}, \bibinfo{person}{Marc Hassenzahl}, \bibinfo{person}{Simon M{\"a}nnlein}, \bibinfo{person}{Melanie Lamara}, \bibinfo{person}{Josef Schumann}, \bibinfo{person}{Kai Eckoldt}, \bibinfo{person}{Matthias Laschke}, {and} \bibinfo{person}{Andreas Butz}.} \bibinfo{year}{2013}\natexlab{}.
\newblock \showarticletitle{Become a member of the last gentlemen: designing for prosocial driving}. In \bibinfo{booktitle}{\emph{Proceedings of the 6th International Conference on Designing Pleasurable Products and Interfaces}}. \bibinfo{pages}{60--66}.
\newblock


\bibitem[Krath et~al\mbox{.}(2021)]%
        {KRATH2021}
\bibfield{author}{\bibinfo{person}{Jeanine Krath}, \bibinfo{person}{Linda Schürmann}, {and} \bibinfo{person}{Harald~F.O. {von Korflesch}}.} \bibinfo{year}{2021}\natexlab{}.
\newblock \showarticletitle{Revealing the theoretical basis of gamification: A systematic review and analysis of theory in research on gamification, serious games and game-based learning}.
\newblock \bibinfo{journal}{\emph{Computers in Human Behavior}}  \bibinfo{volume}{125} (\bibinfo{year}{2021}), \bibinfo{pages}{106963}.
\newblock
\showISSN{0747-5632}
\urldef\tempurl%
\url{https://doi.org/10.1016/j.chb.2021.106963}
\showDOI{\tempurl}


\bibitem[Lee et~al\mbox{.}(2021)]%
        {Lee2021}
\bibfield{author}{\bibinfo{person}{Daegyu Lee}, \bibinfo{person}{Gyuree Kang}, \bibinfo{person}{Boseong Kim}, {and} \bibinfo{person}{D.~Hyunchul Shim}.} \bibinfo{year}{2021}\natexlab{}.
\newblock \showarticletitle{Assistive Delivery Robot Application for Real-World Postal Services}.
\newblock \bibinfo{journal}{\emph{IEEE Access}}  \bibinfo{volume}{9} (\bibinfo{year}{2021}), \bibinfo{pages}{141981--141998}.
\newblock
\urldef\tempurl%
\url{https://doi.org/10.1109/ACCESS.2021.3120618}
\showDOI{\tempurl}


\bibitem[L{\"u}decke et~al\mbox{.}(2021)]%
        {ludecke2021performance}
\bibfield{author}{\bibinfo{person}{Daniel L{\"u}decke}, \bibinfo{person}{Mattan~S Ben-Shachar}, \bibinfo{person}{Indrajeet Patil}, \bibinfo{person}{Philip Waggoner}, {and} \bibinfo{person}{Dominique Makowski}.} \bibinfo{year}{2021}\natexlab{}.
\newblock \showarticletitle{performance: An R package for assessment, comparison and testing of statistical models}.
\newblock \bibinfo{journal}{\emph{Journal of Open Source Software}} \bibinfo{volume}{6}, \bibinfo{number}{60} (\bibinfo{year}{2021}).
\newblock


\bibitem[Ma et~al\mbox{.}(2021)]%
        {Ma2021}
\bibfield{author}{\bibinfo{person}{Shu Ma}, \bibinfo{person}{Wei Zhang}, \bibinfo{person}{Zhen Yang}, \bibinfo{person}{Chunyan Kang}, \bibinfo{person}{Changxu Wu}, \bibinfo{person}{Chunlei Chai}, \bibinfo{person}{Jinlei Shi}, \bibinfo{person}{Yilin Zeng}, {and} \bibinfo{person}{Hongting Li}.} \bibinfo{year}{2021}\natexlab{}.
\newblock \showarticletitle{Take over Gradually in Conditional Automated Driving: The Effect of Two-stage Warning Systems on Situation Awareness, Driving Stress, Takeover Performance, and Acceptance}.
\newblock \bibinfo{journal}{\emph{International Journal of Human–Computer Interaction}} \bibinfo{volume}{37}, \bibinfo{number}{4} (\bibinfo{year}{2021}), \bibinfo{pages}{352--362}.
\newblock
\urldef\tempurl%
\url{https://doi.org/10.1080/10447318.2020.1860514}
\showDOI{\tempurl}


\bibitem[Maiti et~al\mbox{.}(2022)]%
        {maiti22}
\bibfield{author}{\bibinfo{person}{Anindya Maiti}, \bibinfo{person}{Nisha Vinayaga-Sureshkanth}, \bibinfo{person}{Murtuza Jadliwala}, \bibinfo{person}{Raveen Wijewickrama}, {and} \bibinfo{person}{Greg Griffin}.} \bibinfo{year}{2022}\natexlab{}.
\newblock \showarticletitle{Impact of E-Scooters on Pedestrian Safety: A Field Study Using Pedestrian Crowd-Sensing}. In \bibinfo{booktitle}{\emph{2022 IEEE International Conference on Pervasive Computing and Communications Workshops and other Affiliated Events (PerCom Workshops)}}. \bibinfo{pages}{799--805}.
\newblock
\urldef\tempurl%
\url{https://doi.org/10.1109/PerComWorkshops53856.2022.9767450}
\showDOI{\tempurl}


\bibitem[Markkula et~al\mbox{.}(2020)]%
        {Markkula2020}
\bibfield{author}{\bibinfo{person}{G. Markkula}, \bibinfo{person}{R. Madigan}, \bibinfo{person}{D. Nathanael}, \bibinfo{person}{E. Portouli}, \bibinfo{person}{Y.~M. Lee}, \bibinfo{person}{A. Dietrich}, \bibinfo{person}{J. Billington}, \bibinfo{person}{A. Schieben}, {and} \bibinfo{person}{N. Merat}.} \bibinfo{year}{2020}\natexlab{}.
\newblock \showarticletitle{Defining interactions: a conceptual framework for understanding interactive behaviour in human and automated road traffic}.
\newblock \bibinfo{journal}{\emph{Theoretical Issues in Ergonomics Science}} \bibinfo{volume}{21}, \bibinfo{number}{6} (\bibinfo{year}{2020}), \bibinfo{pages}{728--752}.
\newblock
\urldef\tempurl%
\url{https://doi.org/10.1080/1463922X.2020.1736686}
\showDOI{\tempurl}


\bibitem[Mehrotra et~al\mbox{.}(2023)]%
        {mehrotra2023trust}
\bibfield{author}{\bibinfo{person}{Shashank Mehrotra}, \bibinfo{person}{Jacob~G Hunter}, \bibinfo{person}{Matthew Konishi}, \bibinfo{person}{Kumar Akash}, \bibinfo{person}{Zhaobo Zheng}, \bibinfo{person}{Teruhisa Misu}, \bibinfo{person}{Anil Kumar}, \bibinfo{person}{Tahira Reid}, {and} \bibinfo{person}{Neera Jain}.} \bibinfo{year}{2023}\natexlab{}.
\newblock \showarticletitle{Trust in shared automated vehicles: Study on two mobility platforms}.
\newblock \bibinfo{journal}{\emph{arXiv preprint arXiv:2303.09711}} (\bibinfo{year}{2023}).
\newblock


\bibitem[Mehrotra and Roberts(2022)]%
        {mehrotra2022identifying}
\bibfield{author}{\bibinfo{person}{Shashank Mehrotra} {and} \bibinfo{person}{Shannon~C Roberts}.} \bibinfo{year}{2022}\natexlab{}.
\newblock \showarticletitle{Identifying And Improving Young Drivers’ Perceptions Towards Vulnerable Road Users}. In \bibinfo{booktitle}{\emph{Proceedings of the Human Factors and Ergonomics Society Annual Meeting}}, Vol.~\bibinfo{volume}{66}. SAGE Publications Sage CA: Los Angeles, CA, \bibinfo{pages}{1350--1354}.
\newblock


\bibitem[Mehrotra et~al\mbox{.}(2022)]%
        {mehrotra2022human}
\bibfield{author}{\bibinfo{person}{Shashank Mehrotra}, \bibinfo{person}{Meng Wang}, \bibinfo{person}{Nicholas Wong}, \bibinfo{person}{Jah'inaya Parker}, \bibinfo{person}{Shannon~C Roberts}, \bibinfo{person}{Woon Kim}, \bibinfo{person}{Alicia Romo}, {and} \bibinfo{person}{William~J Horrey}.} \bibinfo{year}{2022}\natexlab{}.
\newblock \showarticletitle{Human-machine interfaces and vehicle automation: A review of the literature and recommendations for system design, feedback, and alerts}.
\newblock  (\bibinfo{year}{2022}).
\newblock


\bibitem[Metz(2023)]%
        {forbes_23}
\bibfield{author}{\bibinfo{person}{Jason Metz}.} \bibinfo{year}{2023}\natexlab{}.
\newblock
\newblock
\urldef\tempurl%
\url{https://www.forbes.com/advisor/car-insurance/usage-based-insurance/#:~:text=Usage%2Dbased%20insurance%20programs%20generally,better%20your%20auto%20insurance%20rates.}
\showURL{%
\tempurl}


\bibitem[Narang et~al\mbox{.}(2015)]%
        {narang2015detecting}
\bibfield{author}{\bibinfo{person}{Priya Narang}, \bibinfo{person}{Amar Agarwal}, {and} \bibinfo{person}{Areeckal~S Sanu}.} \bibinfo{year}{2015}\natexlab{}.
\newblock \showarticletitle{Detecting subtle intraocular movements: Enhanced frames per second recording (slow motion) using smartphones}.
\newblock \bibinfo{journal}{\emph{Journal of Cataract \& Refractive Surgery}} \bibinfo{volume}{41}, \bibinfo{number}{6} (\bibinfo{year}{2015}), \bibinfo{pages}{1321--1323}.
\newblock


\bibitem[Nieuwenhuijsen and Khreis(2016)]%
        {nieuwenhuijsen2016car}
\bibfield{author}{\bibinfo{person}{Mark~J Nieuwenhuijsen} {and} \bibinfo{person}{Haneen Khreis}.} \bibinfo{year}{2016}\natexlab{}.
\newblock \showarticletitle{Car free cities: Pathway to healthy urban living}.
\newblock \bibinfo{journal}{\emph{Environment international}}  \bibinfo{volume}{94} (\bibinfo{year}{2016}), \bibinfo{pages}{251--262}.
\newblock


\bibitem[Niu et~al\mbox{.}(2024)]%
        {niu2024beyond}
\bibfield{author}{\bibinfo{person}{Minxue Niu}, \bibinfo{person}{Zhaobo~K Zheng}, \bibinfo{person}{Kumar Akash}, {and} \bibinfo{person}{Teruhisa Misu}.} \bibinfo{year}{2024}\natexlab{}.
\newblock \showarticletitle{Beyond empirical windowing: An attention-based approach for trust prediction in autonomous vehicles}. In \bibinfo{booktitle}{\emph{ICASSP 2024-2024 IEEE International Conference on Acoustics, Speech and Signal Processing (ICASSP)}}. IEEE, \bibinfo{pages}{5615--5619}.
\newblock


\bibitem[Penner. et~al\mbox{.}(2005)]%
        {Penner2005}
\bibfield{author}{\bibinfo{person}{Louis~A. Penner.}, \bibinfo{person}{John~F. Dovidio.}, \bibinfo{person}{Jane~A. Piliavin.}, {and} \bibinfo{person}{David~A. Schroeder.}} \bibinfo{year}{2005}\natexlab{}.
\newblock \showarticletitle{Prosocial Behavior: Multilevel Perspectives}.
\newblock \bibinfo{journal}{\emph{Annual Review of Psychology}} \bibinfo{volume}{56}, \bibinfo{number}{1} (\bibinfo{year}{2005}), \bibinfo{pages}{365--392}.
\newblock
\urldef\tempurl%
\url{https://doi.org/10.1146/annurev.psych.56.091103.070141}
\showDOI{\tempurl}
\newblock
\shownote{PMID: 15709940}.


\bibitem[Petzoldt et~al\mbox{.}(2017)]%
        {petzoldt2017drivers}
\bibfield{author}{\bibinfo{person}{Tibor Petzoldt}, \bibinfo{person}{Katja Schleinitz}, \bibinfo{person}{Josef~F Krems}, {and} \bibinfo{person}{Tina Gehlert}.} \bibinfo{year}{2017}\natexlab{}.
\newblock \showarticletitle{Drivers’ gap acceptance in front of approaching bicycles--Effects of bicycle speed and bicycle type}.
\newblock \bibinfo{journal}{\emph{Safety science}}  \bibinfo{volume}{92} (\bibinfo{year}{2017}), \bibinfo{pages}{283--289}.
\newblock


\bibitem[Pillutla and Chen(1999)]%
        {pillutla1999social}
\bibfield{author}{\bibinfo{person}{Madan~M Pillutla} {and} \bibinfo{person}{Xiao-Ping Chen}.} \bibinfo{year}{1999}\natexlab{}.
\newblock \showarticletitle{Social norms and cooperation in social dilemmas: The effects of context and feedback}.
\newblock \bibinfo{journal}{\emph{Organizational behavior and human decision processes}} \bibinfo{volume}{78}, \bibinfo{number}{2} (\bibinfo{year}{1999}), \bibinfo{pages}{81--103}.
\newblock


\bibitem[Pryer and Bass(1959)]%
        {Preyer1959}
\bibfield{author}{\bibinfo{person}{Margaret~W. Pryer} {and} \bibinfo{person}{Bernard~M. Bass}.} \bibinfo{year}{1959}\natexlab{}.
\newblock \showarticletitle{Some Effects of Feedback on Behavior in Groups}.
\newblock \bibinfo{journal}{\emph{Sociometry}} \bibinfo{volume}{22}, \bibinfo{number}{1} (\bibinfo{year}{1959}), \bibinfo{pages}{56--63}.
\newblock
\showISSN{00380431}
\urldef\tempurl%
\url{http://www.jstor.org/stable/2785612}
\showURL{%
\tempurl}


\bibitem[Shaheen et~al\mbox{.}(2020)]%
        {shaheen2020sharing}
\bibfield{author}{\bibinfo{person}{Susan Shaheen}, \bibinfo{person}{Adam Cohen}, \bibinfo{person}{Nelson Chan}, {and} \bibinfo{person}{Apaar Bansal}.} \bibinfo{year}{2020}\natexlab{}.
\newblock \showarticletitle{Sharing strategies: carsharing, shared micromobility (bikesharing and scooter sharing), transportation network companies, microtransit, and other innovative mobility modes}.
\newblock In \bibinfo{booktitle}{\emph{Transportation, land use, and environmental planning}}. \bibinfo{publisher}{Elsevier}, \bibinfo{pages}{237--262}.
\newblock


\bibitem[Stern et~al\mbox{.}(2019)]%
        {STERN2019}
\bibfield{author}{\bibinfo{person}{Raphael~E. Stern}, \bibinfo{person}{Yuche Chen}, \bibinfo{person}{Miles Churchill}, \bibinfo{person}{Fangyu Wu}, \bibinfo{person}{Maria~Laura {Delle Monache}}, \bibinfo{person}{Benedetto Piccoli}, \bibinfo{person}{Benjamin Seibold}, \bibinfo{person}{Jonathan Sprinkle}, {and} \bibinfo{person}{Daniel~B. Work}.} \bibinfo{year}{2019}\natexlab{}.
\newblock \showarticletitle{Quantifying air quality benefits resulting from few autonomous vehicles stabilizing traffic}.
\newblock \bibinfo{journal}{\emph{Transportation Research Part D: Transport and Environment}}  \bibinfo{volume}{67} (\bibinfo{year}{2019}), \bibinfo{pages}{351--365}.
\newblock
\showISSN{1361-9209}
\urldef\tempurl%
\url{https://doi.org/10.1016/j.trd.2018.12.008}
\showDOI{\tempurl}


\bibitem[Subramanyam et~al\mbox{.}(2023)]%
        {subramanyam2023accident}
\bibfield{author}{\bibinfo{person}{Rohith~Puvvala Subramanyam}, \bibinfo{person}{Abhay Naik}, {and} \bibinfo{person}{Mahima~Agumbe Suresh}.} \bibinfo{year}{2023}\natexlab{}.
\newblock \showarticletitle{Accident Prediction on E-Bikes Using Computer Vision}. In \bibinfo{booktitle}{\emph{2023 IEEE Ninth International Conference on Big Data Computing Service and Applications (BigDataService)}}. IEEE, \bibinfo{pages}{186--190}.
\newblock


\bibitem[Tafidis et~al\mbox{.}(2022)]%
        {Tafidis2022}
\bibfield{author}{\bibinfo{person}{Pavlos Tafidis}, \bibinfo{person}{Haneen Farah}, \bibinfo{person}{Tom Brijs}, {and} \bibinfo{person}{Ali Pirdavani}.} \bibinfo{year}{2022}\natexlab{}.
\newblock \showarticletitle{Safety implications of higher levels of automated vehicles: a scoping review}.
\newblock \bibinfo{journal}{\emph{Transport Reviews}} \bibinfo{volume}{42}, \bibinfo{number}{2} (\bibinfo{year}{2022}), \bibinfo{pages}{245--267}.
\newblock
\urldef\tempurl%
\url{https://doi.org/10.1080/01441647.2021.1971794}
\showDOI{\tempurl}


\bibitem[Thielmann et~al\mbox{.}(2020)]%
        {Thielmann2020}
\bibfield{author}{\bibinfo{person}{Isabel Thielmann}, \bibinfo{person}{Giuliana Spadaro}, {and} \bibinfo{person}{Daniel Balliet}.} \bibinfo{year}{2020}\natexlab{}.
\newblock \showarticletitle{Personality and prosocial behavior: A theoretical framework and meta-analysis}.
\newblock \bibinfo{journal}{\emph{Psychological Bulletin}}  \bibinfo{volume}{146} (\bibinfo{year}{2020}), \bibinfo{pages}{30--90}.
\newblock
Issue 1.
\showISSN{00332909}
\urldef\tempurl%
\url{https://doi.org/10.1037/bul0000217}
\showDOI{\tempurl}


\bibitem[Thorndike(1898)]%
        {Thorndike_1898}
\bibfield{author}{\bibinfo{person}{Edward~L. Thorndike}.} \bibinfo{year}{1898}\natexlab{}.
\newblock \showarticletitle{Animal Intelligence: An experimental study of the associative processes in animals.}
\newblock \bibinfo{journal}{\emph{The Psychological Review: Monograph Supplements}} \bibinfo{volume}{2}, \bibinfo{number}{4} (\bibinfo{year}{1898}), \bibinfo{pages}{i–109}.
\newblock
\urldef\tempurl%
\url{https://doi.org/10.1037/h0092987}
\showDOI{\tempurl}


\bibitem[Tuncer et~al\mbox{.}(2020)]%
        {TUNCER2020}
\bibfield{author}{\bibinfo{person}{Sylvaine Tuncer}, \bibinfo{person}{Eric Laurier}, \bibinfo{person}{Barry Brown}, {and} \bibinfo{person}{Christian Licoppe}.} \bibinfo{year}{2020}\natexlab{}.
\newblock \showarticletitle{Notes on the practices and appearances of e-scooter users in public space}.
\newblock \bibinfo{journal}{\emph{Journal of Transport Geography}}  \bibinfo{volume}{85} (\bibinfo{year}{2020}), \bibinfo{pages}{102702}.
\newblock
\showISSN{0966-6923}
\urldef\tempurl%
\url{https://doi.org/10.1016/j.jtrangeo.2020.102702}
\showDOI{\tempurl}


\bibitem[Wang et~al\mbox{.}(2016)]%
        {wang2016likes}
\bibfield{author}{\bibinfo{person}{Chao Wang}, \bibinfo{person}{Jacques Terken}, \bibinfo{person}{Jun Hu}, {and} \bibinfo{person}{Matthias Rauterberg}.} \bibinfo{year}{2016}\natexlab{}.
\newblock \showarticletitle{" Likes" and" Dislikes" on the Road: A Social Feedback System for Improving Driving Behavior}. In \bibinfo{booktitle}{\emph{Proceedings of the 8th International Conference on Automotive User Interfaces and Interactive Vehicular Applications}}. \bibinfo{pages}{43--50}.
\newblock


\bibitem[Ward et~al\mbox{.}(2020)]%
        {ward2020}
\bibfield{author}{\bibinfo{person}{Nicholas Ward}, \bibinfo{person}{Kari Finley}, \bibinfo{person}{Jay Otto}, \bibinfo{person}{David Kack}, \bibinfo{person}{Rebecca Gleason}, {and} \bibinfo{person}{Taylor Lonsdale}.} \bibinfo{year}{2020}\natexlab{}.
\newblock \showarticletitle{Traffic safety culture and prosocial driver behavior for safer vehicle-bicyclist interactions}.
\newblock \bibinfo{journal}{\emph{Journal of Safety Research}}  \bibinfo{volume}{75} (\bibinfo{year}{2020}), \bibinfo{pages}{24--31}.
\newblock
\urldef\tempurl%
\url{https://doi.org/10.1016/j.jsr.2020.07.003}
\showDOI{\tempurl}


\bibitem[Weinberg et~al\mbox{.}(2023)]%
        {weinberg2023}
\bibfield{author}{\bibinfo{person}{David Weinberg}, \bibinfo{person}{Healy Dwyer}, \bibinfo{person}{Sarah~E. Fox}, {and} \bibinfo{person}{Nikolas Martelaro}.} \bibinfo{year}{2023}\natexlab{}.
\newblock \showarticletitle{Sharing the Sidewalk: Observing Delivery Robot Interactions with Pedestrians during a Pilot in Pittsburgh, PA}.
\newblock \bibinfo{journal}{\emph{Multimodal Technologies and Interaction}} \bibinfo{volume}{7}, \bibinfo{number}{5} (\bibinfo{year}{2023}).
\newblock
\showISSN{2414-4088}
\urldef\tempurl%
\url{https://doi.org/10.3390/mti7050053}
\showDOI{\tempurl}


\bibitem[Whitson(2013)]%
        {whitson2013gaming}
\bibfield{author}{\bibinfo{person}{Jennifer~R Whitson}.} \bibinfo{year}{2013}\natexlab{}.
\newblock \showarticletitle{Gaming the quantified self}.
\newblock \bibinfo{journal}{\emph{Surveillance \& Society}} \bibinfo{volume}{11}, \bibinfo{number}{1/2} (\bibinfo{year}{2013}), \bibinfo{pages}{163--176}.
\newblock


\bibitem[Winton(2022)]%
        {forbes2022}
\bibfield{author}{\bibinfo{person}{Neil Winton}.} \bibinfo{year}{2022}\natexlab{}.
\newblock \showarticletitle{Computer Driven Autos Still Years Away Despite Massive Investment}.
\newblock \bibinfo{journal}{\emph{Forbes}} (\bibinfo{year}{2022}).
\newblock
\urldef\tempurl%
\url{https://www.forbes.com/sites/neilwinton/2022/02/27/computer-driven-autos-still-years-away-despite-massive-investment/?sh=74dc936618cc}
\showURL{%
\tempurl}


\bibitem[Xu et~al\mbox{.}(2023)]%
        {xu2023examining}
\bibfield{author}{\bibinfo{person}{Wei Xu}, \bibinfo{person}{Le-Ying Yang}, \bibinfo{person}{Xiao Liu}, {and} \bibinfo{person}{Pin-Nv Jin}.} \bibinfo{year}{2023}\natexlab{}.
\newblock \showarticletitle{Examining the effects of different forms of teacher feedback intervention for learners' cognitive and emotional interaction in online collaborative discussion: A visualization method for process mining based on text automatic analysis}.
\newblock \bibinfo{journal}{\emph{Education and Information Technologies}} (\bibinfo{year}{2023}), \bibinfo{pages}{1--27}.
\newblock


\bibitem[Zheng et~al\mbox{.}(2022)]%
        {zheng2022identification}
\bibfield{author}{\bibinfo{person}{Zhaobo Zheng}, \bibinfo{person}{Kumar Akash}, \bibinfo{person}{Teruhisa Misu}, \bibinfo{person}{Vidya Krishnamoorthy}, \bibinfo{person}{Miaomiao Dong}, \bibinfo{person}{Yuni Lee}, {and} \bibinfo{person}{Gaojian Huang}.} \bibinfo{year}{2022}\natexlab{}.
\newblock \showarticletitle{Identification of adaptive driving style preference through implicit inputs in sae l2 vehicles}. In \bibinfo{booktitle}{\emph{Proceedings of the 2022 International Conference on Multimodal Interaction}}. \bibinfo{pages}{468--475}.
\newblock


\bibitem[Şengül and Mostofi(2021)]%
        {sengul21}
\bibfield{author}{\bibinfo{person}{Buket Şengül} {and} \bibinfo{person}{Hamid Mostofi}.} \bibinfo{year}{2021}\natexlab{}.
\newblock \showarticletitle{Impacts of E-Micromobility on the Sustainability of Urban Transportation—A Systematic Review}.
\newblock \bibinfo{journal}{\emph{Applied Sciences}} \bibinfo{volume}{11}, \bibinfo{number}{13} (\bibinfo{year}{2021}).
\newblock
\showISSN{2076-3417}
\urldef\tempurl%
\url{https://doi.org/10.3390/app11135851}
\showDOI{\tempurl}


\end{thebibliography}
\end{document}